\documentclass[useAMS,usenatbib]{mn2e}

\usepackage{graphicx}
\newcommand{\ltsimeq}{\raisebox{-0.6ex}{$\,\stackrel
        {\raisebox{-.2ex}{$\textstyle <$}}{\sim}\,$}}

\title[2:1 commensurability in a gas giant -- Super-Earth system]
{Occurrence of the 2:1 commensurability in a gas giant -- Super-Earth system}
\author[E. Podlewska-Gaca and E. Szuszkiewicz]
{E. Podlewska-Gaca$$\thanks{E-mail: edytap@univ.szczecin.pl (EP)} and 
E. Szuszkiewicz$$\thanks{E-mail: szusz@fermi.fiz.univ.szczecin.pl (ES)} \\
Institute of Physics and CASA*, University of Szczecin, ul. Wielkopolska 15,
70-451 Szczecin, Poland}

\begin{document}

\date{Accepted; Received; in original form }

\pagerange{\pageref{firstpage}--\pageref{lastpage}} \pubyear{2010}

\maketitle

\label{firstpage}

\begin{abstract}
We investigate how the conditions occurring in a protoplanetary disc may
determine the final structure of a planetary system  emerging from such
a disc. We concentrate our attention on the dynamical interactions 
between 
disc and planets leading to orbital migration, which in turn, in 
favourable circumstances, can drive planets into a mean-motion 
commensurability.  We find that for a system containing a gas giant on 
the external orbit and a Super-Earth on the internal one, both
 embedded in a gaseous disc, the 2:1 resonance is a very 
likely configuration, so one can expect it as an outcome of the early 
phases of the planetary system formation. Our conclusion is based on 
an extensive computational survey in which we ask what are the disc 
properties (the surface density and the viscosity) for which the 2:1 
commensurability may be attained. To answer this question we employ a 
full two-dimensional hydrodynamic treatment of the disc-planet 
interactions.  In general terms, we can claim that the 
conditions which favour the 2:1 mean-motion resonance exist in the 
protoplanetary discs with mass accretion rates
in the range of 5$\cdot 10^{-9} - 5\cdot 10^{-8}$ M$_{\odot}$/year.
For accretion rates higher than those needed
for the 2:1 commensurability we observe a variety of behaviours, among
them  the passage to the 3:2 resonance, the scattering 
of the Super-Earth   
or the divergent migration caused by the outward migration 
of the gas giant.
The results we have obtained from numerical simulations are compared 
with the predictions coming from the existing analytical expressions
of the migration speed and the strength of the mean motion resonances.
The conditions that we have found for the attainment of the 2:1 
commensurability are discussed in the framework of the properties 
of protoplanetary discs that are known from the observations.

\end{abstract}


\begin{figure*}
\begin{minipage}{180mm}
\centering
\includegraphics[width=120mm]{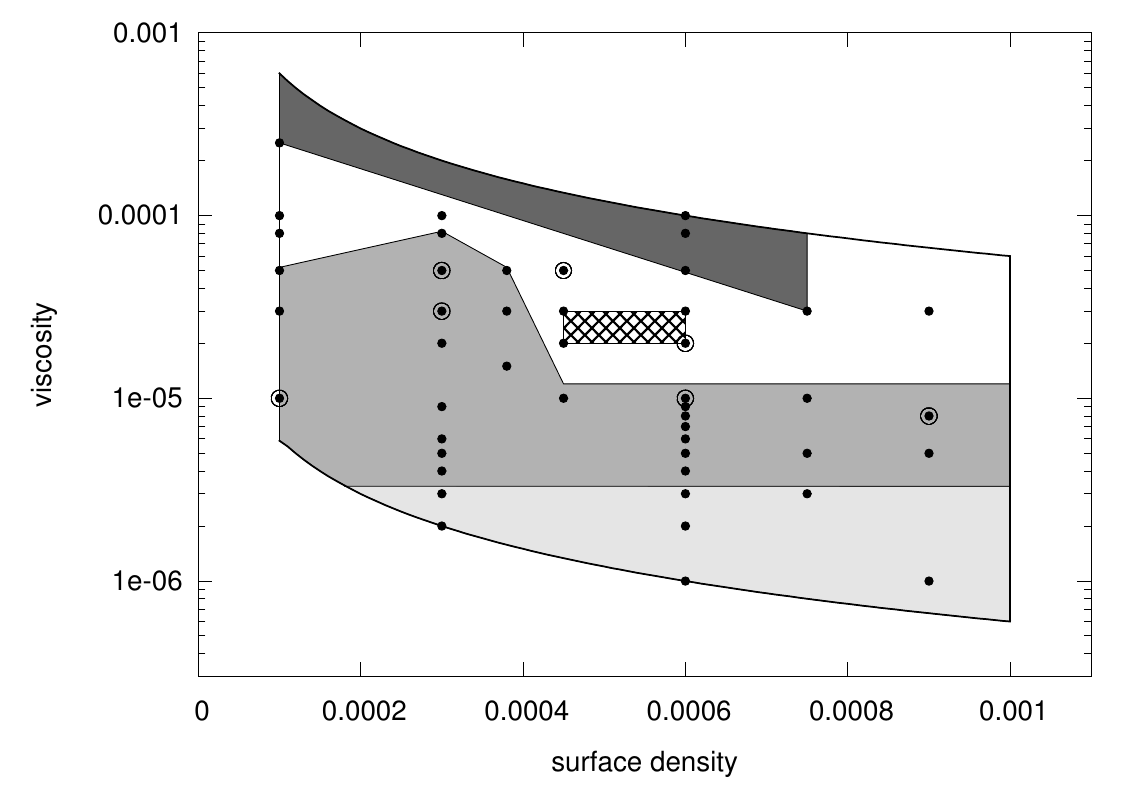}
\caption{\label{fig1}{The parameter space of the properties of the 
protoplanetary discs: surface density - viscosity plane. 
 The central medium gray part denotes the region where planets attain the 2:1
commensurability and the small hatched area favours the formation of the 3:2 resonance.
Upper dark gray and lower light gray colours show the parameters for which
the migration is divergent, and the white colour denotes the transition
region where the variety of behaviours have been observed.
}}
\end{minipage}
\end{figure*}

\section{Introduction}
\label{introduction}
Mean-motion resonances are ubiquitous phenomena in  planetary systems. 
They carry  important information about the formation processes and
the further evolution of those systems. Moreover, their occurrence may 
increase the  chance of detecting planets which happen to be in a resonant 
configuration \citep{agol2005, holman}. They might also be  able to tell us 
more about the properties 
of the protoplanetary discs in which  planets are born. 
The increasing number of low-mass planets being discovered in  multiple
planetary systems \citep{mayor9a, mayor9b, lovis10}
is the first strong observational motivation
for  the studies of planetary formation and evolution.
Another motivation is connected with the growing amount of data
coming from  high angular resolution observations of protoplanetary
discs \citep{Andrews09, hughes, isella, olo, huelamo}.
The excitement in this area of research is stimulated also by the fast
progress in theoretical investigations of the orbital migration due to
the action of tidal torques \citep{linpap93, ward97, tanaka02, paapap08, paapap09}.
The convergent migration of planets or
protoplanetary cores is  one of the most promising processes to
explain the formation
of resonant configurations.  Extensive studies of  possible resonant
structures have been performed among others by \cite{nelpap2002}
and \cite{papszusz2005,papszusz2010}. In  \cite{nelpap2002}
the possible commensurabilities to be
expected in the system of planets in the Jovian mass range
are investigated,  while in
\cite{papszusz2005,papszusz2010}  migrating planets
in the terrestrial mass range are considered and the  conditions for
the occurrence
of first order mean-motion resonances are provided.
\cite{
fonel07b}, \cite{zhou}, \cite{raymond06}, \cite{mandell} have shown that
terrestrial planets can grow and be retained in the hot-Jupiter systems
and that their resonant captures by a gas giant are quite common.
In the present paper we consider a system containing a Jupiter-like
planet evolving together with a Super-Earth in a gaseous disc. We perform
a survey of the different disc properties in which this system is
embedded using full hydrodynamical calculations to simulate the dynamical
interactions between discs and planets.
Our purpose here is to find out what are the conditions for the occurrence
of the 2:1 commensurability in such systems  and at the same time 
to get an insight into
the environment suitable for the formation of planets on the basis of the 
existence of certain mean-motion commensurabilities in  young  planetary 
systems.
It has been shown in \cite{paperI, paperII} 
that in the gas giant -- Super-Earth systems the planets embedded in a 
gaseous disc are easily locked in the commensurabilities if and
only if the Super-Earth 
is located inside the orbit of the gas giant. For this reason, 
we consider here exactly that kind of configurations in which
 the Super-Earth is the
inner and the gas giant is the outer planet of the system.
 In the previous paper 
\citep{paperI}, where the same system has been studied, 
the 2:1 commensurability was not included because of computational 
limitations. 

This paper is organized as follows: In Section \ref{survey} we describe 
how our computational survey has been performed. Next, we illustrate 
the relevant features of the evolution of Jupiter-like planets in 
discs with different viscosities. In Section \ref{results} we give the 
results of our survey presenting different configurations achieved during 
the tidal evolution of the planets in the disc.  These results are
compared in Section \ref{theoretical} with the theoretical expectation 
based on the analytical and
semi-analytical expressions for the migration speed of the planets and
the strength of the mean motion resonances.
In Section \ref{observational}  we point those observed 
protoplanetary discs in which the physical conditions that favour
the occurrence of the 2:1 resonance are present.
Finally, our findings are summarized and discussed in 
Section \ref{conclusions}.   

\section{Setting the scene for the survey}
\label{survey}
In order to determine  possible resonances induced by  planet
migration in the presence of a gaseous disc in a system containing
a Super-Earth and a gas giant, we have performed a computational survey
by means of full two-dimensional hydrodynamic simulations.
The mass of the Super-Earth is taken to be 5.5 M$_{\oplus}$ and that of
the gas giant 1 M$_J$ (here M$_{\oplus}$ and M$_J$ denote the masses 
of the Earth
and the Jupiter respectively). This pair of planets is used in our
simulations as a probe of the disc properties which favour  
the resonant capture
of the Super-Earth by the gas giant. We have fixed the disc aspect ratio
to $h=0.05$, which is a typical value for protoplanetary discs. 
Herewith, we have assumed for the initial
surface density the classical power law 
parameterization
\begin{equation}
\Sigma (r) = \Sigma _0 \left( {r \over 5.2 AU}\right)^{-p}
\label{sigma}
\end{equation}
with $p=0$.
We will comment on these
assumptions in Sections \ref{theoretical} and \ref{observational}.
In our survey different discs will be analyzed, whose properties will
be distinguished by using two parameters, namely the surface density
value $\Sigma$ and the coefficient of kinematic viscosity $\nu$, which is 
constant in space and time.
In the following, these two parameters will be referred to as 
``disc properties''.
The parameter space for our studies is illustrated in Fig. \ref{fig1}.
The value of the surface density $\Sigma$ varies 
from $10^{-4}$ to $10^{-3}$
in the dimensionless units used in the code. 
The unit of mass is the mass of the central star $M$ (here we take  
$M=M_{\odot}$) and the 
unit of length $r_p$ is the initial
position of the inner planet (the Super-Earth in our case).
The unit of time in the code is given by $(GM/r_p^
3)^{−1/2}$ ($G$ is the gravitational constant).
This quantity amounts to $(1/2\pi)$
times the orbital period of the initial orbit of the Super-Earth.
The value of $6\cdot 10^{-4}$ of the surface density
in these units is consistent with the Minimum Mass Solar Nebula 
(MMSN)  \citep{hayashi} at the present position of Jupiter in our
planetary system,  i.e. 5.2 AU. The range of the surface 
densities considered
here expressed in physical units is the following: 
30 - 300 $\rm g/cm^2$. 
In order to constrain the values of the kinematic viscosity $\nu$ 
for our studies, we have used the relation between the kinematic
viscosity and the accretion rate $\dot M$ observed in  T Tauri 
stars,  which are the best laboratories for studying the early stages of
planetary evolution. For a stationary thin accretion disc this relation reads 
\begin{equation}
\Sigma \nu = {\dot M \over 3\pi}.
\label{signu}
\end{equation}
The upper and lower
curves in Fig. \ref{fig1} are  the constant accretion
rate levels. For the upper curve we have adopted
$3\cdot 10^{-7}$ ${M_{\odot} /
\rm {year}}$  and for the lower one
$3\cdot 10^{-9}$ ${M_{\odot} /
\rm {year}}$, which are typical accretion rates observed in T Tauri 
stars. The viscosities obtained in this way vary from $6\cdot 10^{-4}$ to
$6\cdot 10^{-7}$ in terms of the inverse of the Reynolds number
at the initial location of the Super-Earth.    
Those two curves, together with the minimal and maximal values
of the surface density, form the region of parameter space which is of
interest for our studies.
In the region defined in this way we have
chosen 47 initial disc configurations for which we have performed
numerical simulations.
Each black dot in Fig. \ref{fig1} represents
a protoplanetary disc of given
surface density and viscosity. The circles around some of the black dots
indicate those discs which are of particular interest and are described
in detail in the next Sections.
The sampling of the parameter space is not uniform. We have
calculated the evolution of planets only for those discs which were relevant
for our purpose, namely for determining the variety of the final
architectures of the planets.

We have performed full two-dimensional hydrodynamic calculations using the 
grid based code NIRVANA. For details on the numerical scheme and 
the adopted code
see \cite{nelson2000}. The planets interact with each other and with 
the protoplanetary disc in which they are embedded. These dynamical 
interactions take place in the central
potential of a solar mass host star. The disc is locally isothermal.
It extends from 0.33 to 4 in  dimensionless units. We use the resolution 
of 384 x 512 grid cells in the radial and the azimuthal directions respectively.
 The grid spacing
in both coordinate directions is uniform.
The computational domain is
a  ring, in which the azimuthal coordinate $\phi$ 
takes its values in
the interval [0, 2$\pi$].
We choose open boundary conditions in the radial direction  and periodic in 
the azimuthal one.
We do not exclude any material from the Hill sphere and we discuss the
consequences of this in Section \ref{theoretical}.
The potential is softened with the parameter $b = 0.8 H$, where $H$ is
the semi-thickness of the disc at the planet position.
The initial eccentricity of both planets is set to zero.

\begin{figure}
\centering
\includegraphics[width=85mm]{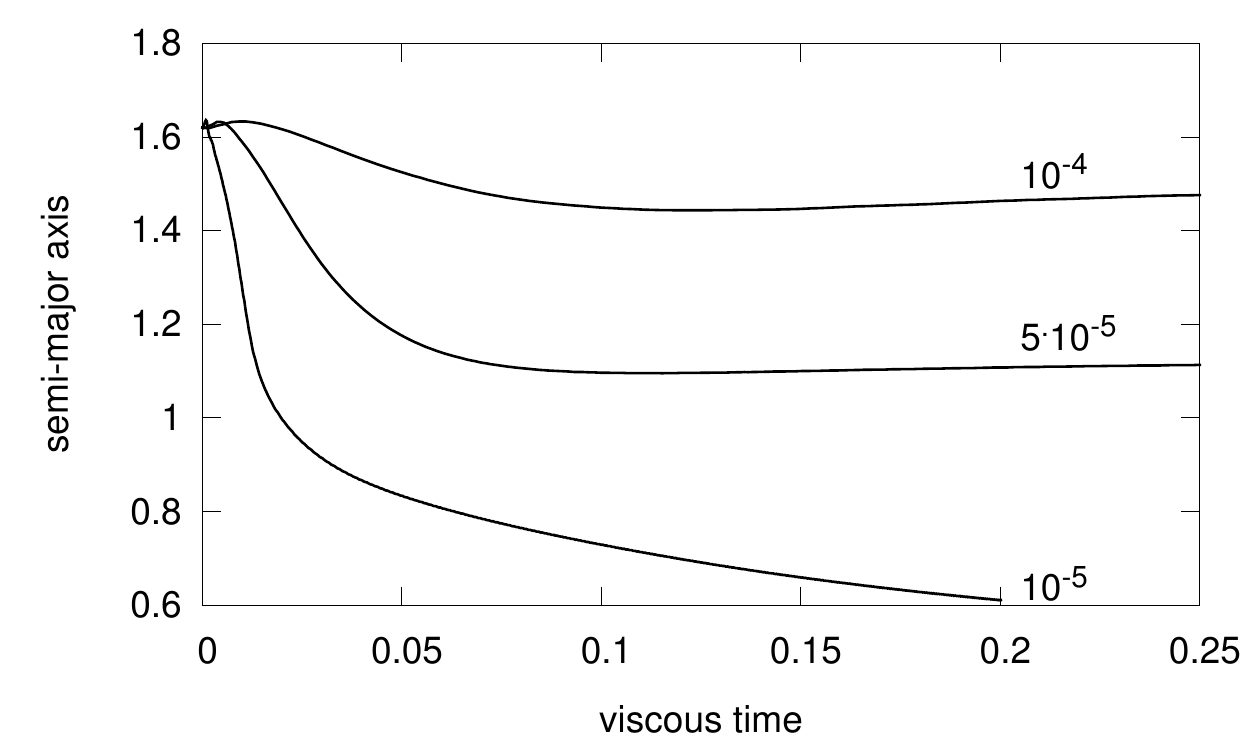}
\caption{\label{fig2}{The semi-major axis evolution of the migrating 
Jupiter-like planet in the disc with  different values of the viscosity 
and with the surface density corresponding to the MMSN at 5.2 AU.
The curves are labeled by the value of the viscosity in the disc
given in terms of the inverse of the Reynolds number. The 'viscous time' 
means the time expressed in viscous time units.
}}
\end{figure}

\section{Gas giant migrating in discs with different viscosities}
\label{jupiter}
It has been established that the migration of the gas giant in  
protoplanetary discs depends on the value of the viscosity
\citep{linpap86, linpap93}. The
viscosity controls the processes responsible for the gap opening 
in the disc which is crucial for further evolution 
of the orbit 
of the planet. 
We have checked that for viscosities lower than $10^{-5}$ 
and typical surface densities, the migration 
of the Jupiter-like
planet in a reasonably good approximation, proceeds on a viscous timescale
as in the classical Type II regime.

\begin{figure}
\centering
\includegraphics[width=85mm]{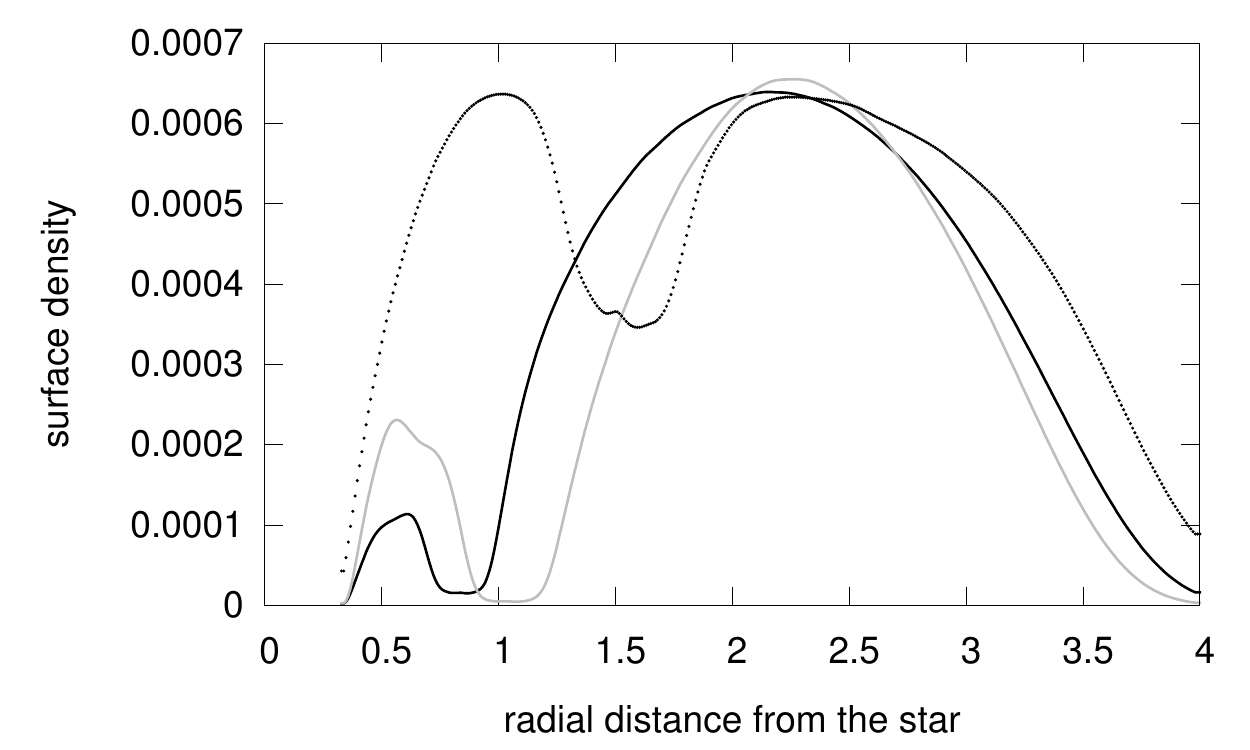}
\caption{\label{fig12} The surface density profiles at 0.05
$t_{\nu}$  for three different viscosities: $5\cdot 10^{-6}$ 
(grey line), $10^{-5}$ (black line)
and $10^{-4}$ (dots). 
}
\end{figure}

Relatively recently, 
\cite{crimor2007} found that for  sufficiently high values of the viscosity,
corresponding to Reynolds numbers of around $10^{4}$, the migration of 
the gas giant tends to change its direction and the planet moves outwards. 
The values of viscosity 
for which they have obtained the outward migration
of the Jupiter-like planet
 are present also in our parameter space so, 
we have been able to
verify if the gas giant behaves similarly 
in the discs analyzed in our survey.
The results, illustrated in Fig. \ref{fig2},  show the evolution of 
the semi-major axis of the Jupiter
mass planet placed in the protoplanetary disc with the surface density
equal to the MMSN at 5.2 AU and  different kinematic
viscosities. Following \cite{crimor2007}, in Fig. \ref{fig2} and
in all other figures where time is present, the
time is measured in
units of the viscous time $t_{\nu}$, which is defined 
in the following way
\begin{equation}
t_{\nu} = {r_p^2 \over \nu}.
\end{equation}
If the viscosity becomes higher 
than $10^{-5}$, 
the Jupiter migration slows down, and if the viscosity reaches 
the value of 
$5\cdot 10^{-5}$ 
or higher,  the migration of the Jupiter is reversed.  
It means that the Jupiter-like planet
will not be subjected to proper type II migration for sufficiently
high disc viscosities.
Our results are thus consistent with those 
obtained by \cite{crimor2007}  despite the differences in the 
initial surface density profile and the numerical treatment adopted
in this paper. To complete the picture of a gas giant migrating 
in the discs with different viscosities,  
we present also the surface density profiles (Fig. \ref{fig12}) 
for three of our simulations characterized by viscosity
values of $5\cdot 10^{-6}$, 
$10^{-5}$ and $10^{-4}$ respectively. All profiles are shown at the same
time of the simulations, namely  
at 0.05 $t_{\nu}$.
This figure  clearly illustrates how the gap profile  depends on
the value of the viscosity.  It is also 
worth mentioning that the mass of the
inner disc critically depends on the position of the gas giant with
respect to the inner edge of the grid as  has
been shown by \cite{crimor2007}.
The outward migration of the Jupiter has  an important consequence 
for the resonant
captures described in Section \ref{results}. 
\begin{figure*}
\begin{minipage}{180mm}
\centering
\includegraphics[width=85mm]{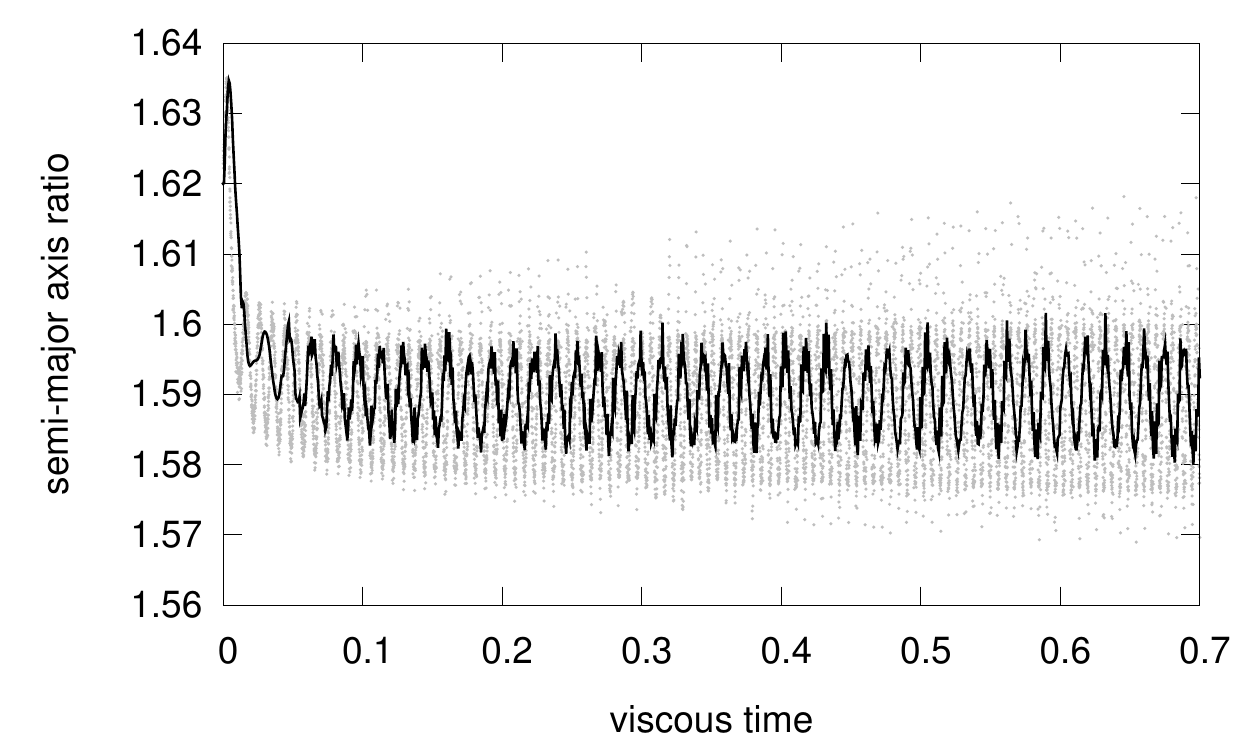}
\includegraphics[width=85mm]{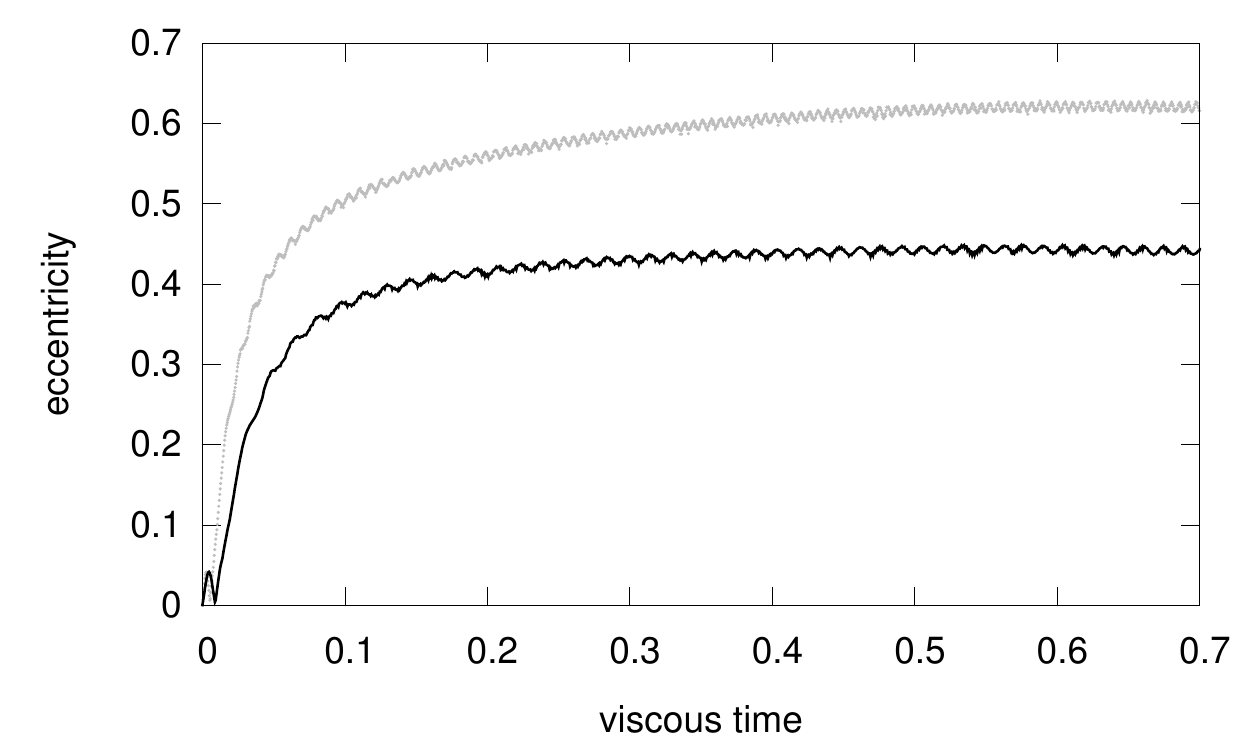}
\caption{\label{fig3}{The evolution of the  semi-major axis ratios
(left panel),
the Super-Earth
eccentricities (right panel)
due to the disc-planet interaction for two discs with the same surface
density equal to $3\cdot 10^{-4}$ and
two values of the kinematic viscosity, namely $5\cdot 10^{-5}$ (black) and
$3\cdot 10^{-5}$ (grey).
}}
\end{minipage}
\end{figure*}

\begin{figure*}
\begin{minipage}{180mm}
\centering
\includegraphics[width=85mm]{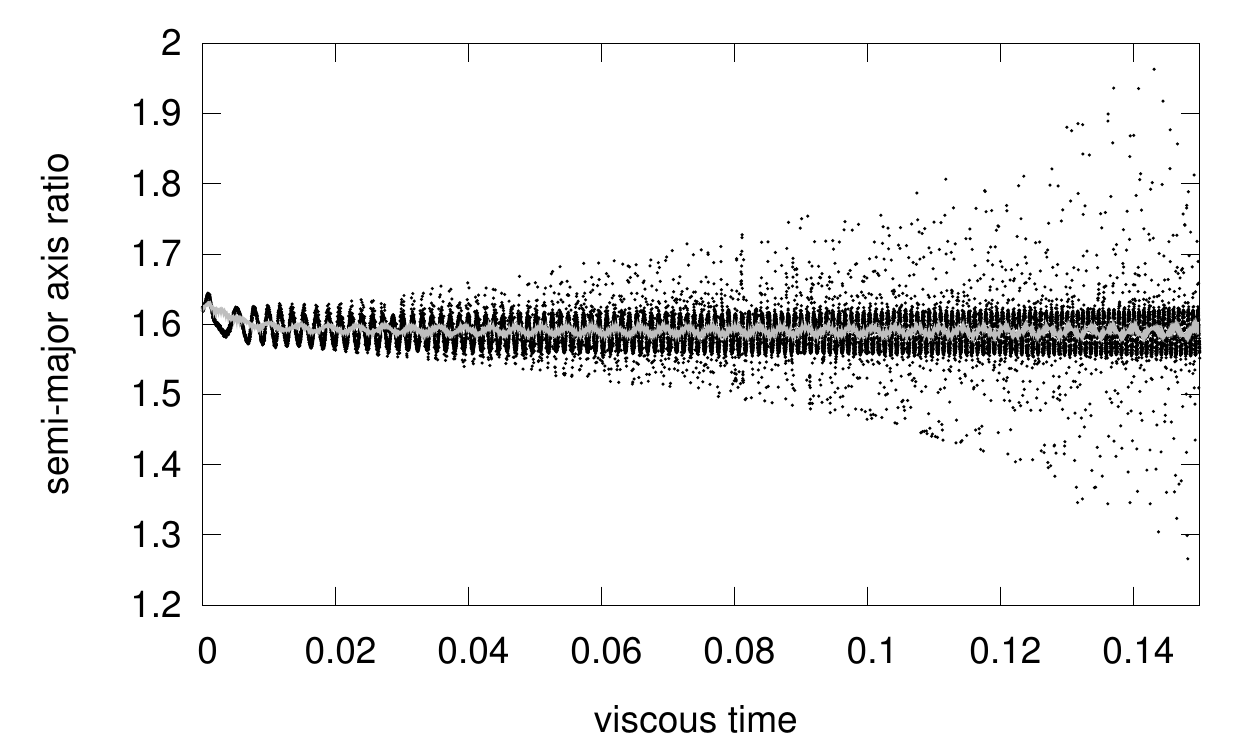}
\includegraphics[width=85mm]{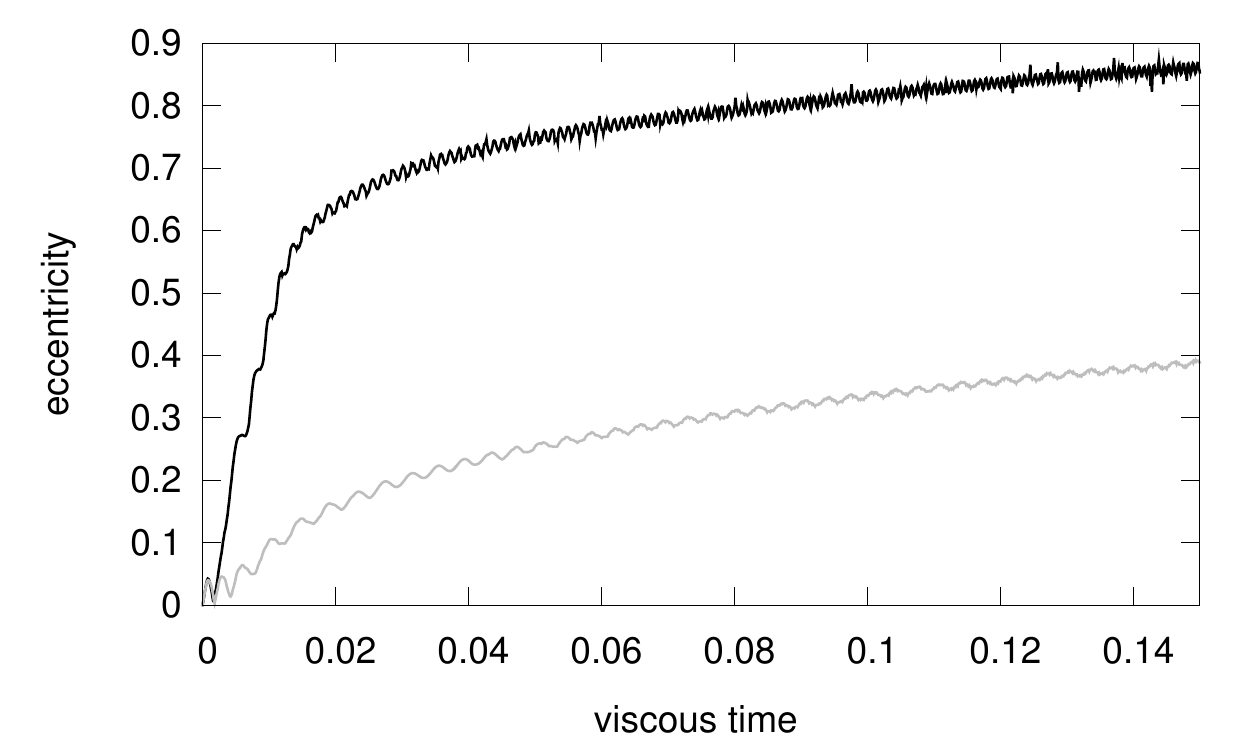}
\caption{\label{fig4}{
The evolution of the  semi-major axis ratios
(left panel),
the Super-Earth
eccentricities (right panel)
due to the disc-planet interaction for two discs with the same viscosity
equal to $10^{-5}$ and
two values of the surface density, namely $6\cdot 10^{-4}$ (black) and
$10^{-4}$ (grey).
}}
\end{minipage}
\end{figure*}

\section{Expected commensurabilities 
for a variety of protoplanetary disc properties}
\label{results}
The planets have been placed in various discs with surface densities and 
viscosities taken from the range shown in Fig. \ref{fig1} and described 
in Section \ref{survey}. 
The initial separation between the  planets has been chosen to be slightly 
larger than that required for the 2:1 resonance. Next, we have followed 
the evolution
of this system for roughly $2\cdot 10^3$ orbits  of the inner planet
i.e. the Super-Earth.
Depending on the final configuration of the planets, we have divided the whole
parameter space into five regions. The medium grey part of the diagram 
in Fig. \ref{fig1} is characterized by the fact that the most
likely outcome of the evolution of the migrating planets is the resonant
capture in the 2:1 commensurability. The disc with properties lying in
the small central hatched region  favours the formation of the
3:2 resonance. For discs with viscosity smaller than $4\cdot 10^{-6}$
the differential 
migration is divergent and we do not expect to find any commensurability
in this case.
The same is true for discs with surface densities and viscosities taken
from the dark grey part of the diagram in Fig. \ref{fig1}. In this region 
the migration of the Jupiter is
directed outward, while the Super-Earth migrates inward, so that the resonant 
capture
is not possible. A transition region in which a variety
of different  behaviours has been observed is denoted in white.

At this point, we would like to describe briefly the typical behaviour of 
the system 
in each
region of the parameter space shown in Fig. \ref{fig1}.

\subsection{The 2:1 commensurability}
\label{2:1}
The capture in the 2:1 resonance is a very likely outcome of the planet
evolution in the disc. Indeed, the medium dark grey region of the diagram
in Fig. \ref{fig1}, 
in which the physical conditions in the disc favour
the  achievement of just that commensurability, is the most pronounced one. 
As it turns out from Fig. \ref{fig1},
the best conditions for the occurrence of the 2:1 resonance exist in the
region in which  the surface densities are  relatively low 
(from $10^{-4}$ till 
3.5$\cdot 10^{-4}$) 
and the viscosity ranges from
4$\cdot 10^{-6}$ till almost $10^{-4}$. For higher surface
densities, in the range from 4.5$\cdot 10^{-4}$ till $10^{-3}$, the
spread of viscosities useful for attaining the 2:1 resonance
shrinks to the interval from 4$\cdot 10^{-6}$
till $10^{-5}$.

First, we investigate how the planets locked in the 2:1 resonance
evolve in time  in discs which differ among themselves only
by the value of the viscosity, while the surface density is kept fixed. 
For this purpose we choose 
$\Sigma = 3\cdot 10^{-4}$ because, as it is shown in Fig. \ref{fig1}, 
this value of the surface density
gives us the biggest range of viscosities 
for which the resonance under consideration is possible.
From our analysis it turns out that
the semi-major axis ratios oscillate always
around the position of the 
resonance and we can identify
two characteristic behaviours of this oscillation. The first
kind of behaviour appears if the 
viscosity is $3\cdot 10^{-5}$ or lower. In that case 
the semi-major axis ratios 
oscillate with
a large amplitude, the same for all discs. The second
kind of behaviour occurs if the viscosity is 
higher than $3\cdot 10^{-5}$ then the amplitude of the  
oscillations decreases together with increasing viscosity. 
Examples of these two behaviours are displayed 
in Fig. \ref{fig3} (left panel)  
for
two values of the kinematic viscosity, namely $5\cdot 10^{-5}$ (black) and
$3\cdot 10^{-5}$ (grey).
 The amplitude of the oscillations
is a straightforward consequence
of the excitation of the Super-Earth eccentricity, shown 
in Fig. \ref{fig3} (right panel).
Let us also notice that
after the capture in the mean motion commensurability the resonant 
interactions drive the eccentricities of the Super-Earth until 
their action is balanced by
the circularization
due to disc tides. For all  discs with 
viscosities $3\cdot 10^{-5}$ or lower  we have observed 
that the eccentricities 
of the Super-Earth  
increase until they attain an equilibrium value of about 0.62. 
An example is 
given in 
 Fig. \ref{fig3} (right panel, 
grey line). For discs with very high viscosities
there is still a lot of material in the vicinity of the Super-Earth,
so the circularization forces act more efficiently and the final
eccentricity (black curve in the right panel of 
Fig. \ref{fig3}) grows to a value
lower than 0.62.
Thus, there is a correlation between 
the final eccentricity
and the surface density of the disc around the position of the Super-Earth.
In the case of higher surface density the eccentricity is lower.

Next,  we perform a similar investigation but this time we fix the
value of the viscosity in the disc and check how 
the evolution of planets locked in the 2:1 resonance
depends on the initial surface density of the disc (or
equivalently on disc mass).  
The value of viscosity is taken to be $\nu = 10^{-5}$, which is
a typical value for protoplanetary discs. 
In Fig. \ref{fig4} we show the evolution of the semi-major axis 
ratios and the
eccentricities of the Super-Earth for  discs with this typical viscosity
and two values of the surface density, namely $\Sigma = 10^{-4}$ (grey
colour) and $\Sigma = 6\cdot 10^{-4}$ (black colour).
The value of the oscillation amplitude of the semi-major axis 
is again the consequence
of the eccentricity evolution. 
The eccentricity of the 
Super-Earth in the 2:1 resonance
is  bigger if the 
initial differential migration of the planets is faster  which is in
agreement with the Eq. (16) of \cite{crida08}.
It can be seen from Fig. \ref{fig4} that  
both presented there eccentricities  do not reach an equilibrium value.
For the disc with higher surface density  
($\Sigma = 6\cdot 10^{-4}$) and viscosity $\nu=10^{-5}$ it occurs,
because 
the Super-Earth arrives at the inner edge of the disc before the
eccentricity attains an equilibrium value. Instead, 
for the disc with lower surface density 
($\Sigma = 10^{-4}$) and the same viscosity
there is not enough material around the position of the
Super-Earth in order to balance the resonant interactions. 
This is why the eccentricity continues to grow till 
the end of the run.

\begin{figure}
\centering
\includegraphics[width=85mm]{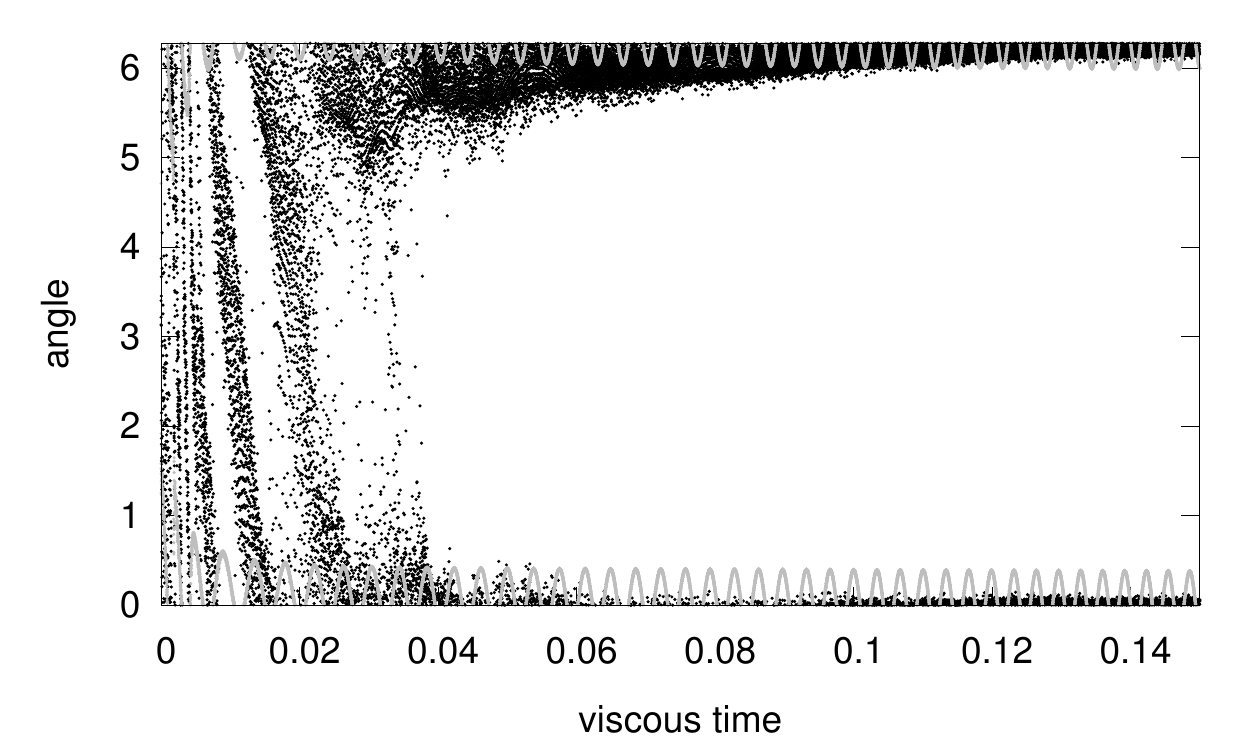}
\caption{\label{fig5}{
The evolution of the angles between the apsidal lines (black colour)
and the resonant angle
(grey colour)
for the disc with the kinematic viscosity $10^{-5}$ and the surface
density equals to $10^{-4}$. 
}}
\end{figure}

In all simulations in which the 2:1 resonance has been observed
the angles between 
the apsidal lines and the resonant angles librate around zero.
In Fig. \ref{fig5} we plot the evolution of the angle between the apsidal
lines (black colour) and the resonant angle, defined as  
$ 2\lambda_J-\lambda_{s}-\omega_{s}$ (grey colour) for 
$\Sigma = 10^{-4}$  and $\nu = 10^{-5}$. Here, $\lambda_J$ and
$\lambda_{s}$
are the mean longitudes of the Jupiter and the Super-Earth
respectively. 
The longitude of the pericentre of the Super-Earth is denoted by
$\omega_{s}$.

Our calculations have been able to determine also the lower limit 
of the viscosity necessary for the
occurrence of the 2:1 resonance. We have noticed that  below $\nu =
4\cdot 10^{-6}$ there is no possibility of attaining the commensurability 
because the differential migration becomes divergent.
\begin{figure*}
\begin{minipage}{180mm}
\centering
\includegraphics[width=85mm]{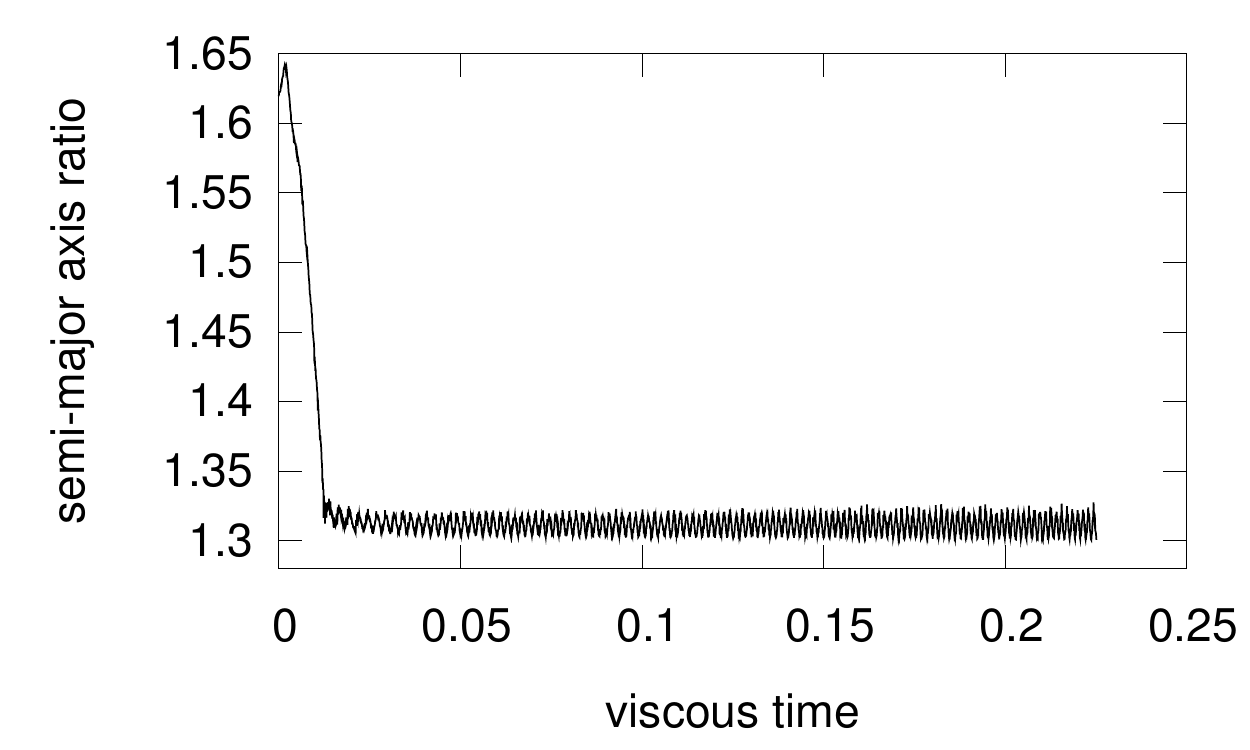}
\includegraphics[width=85mm]{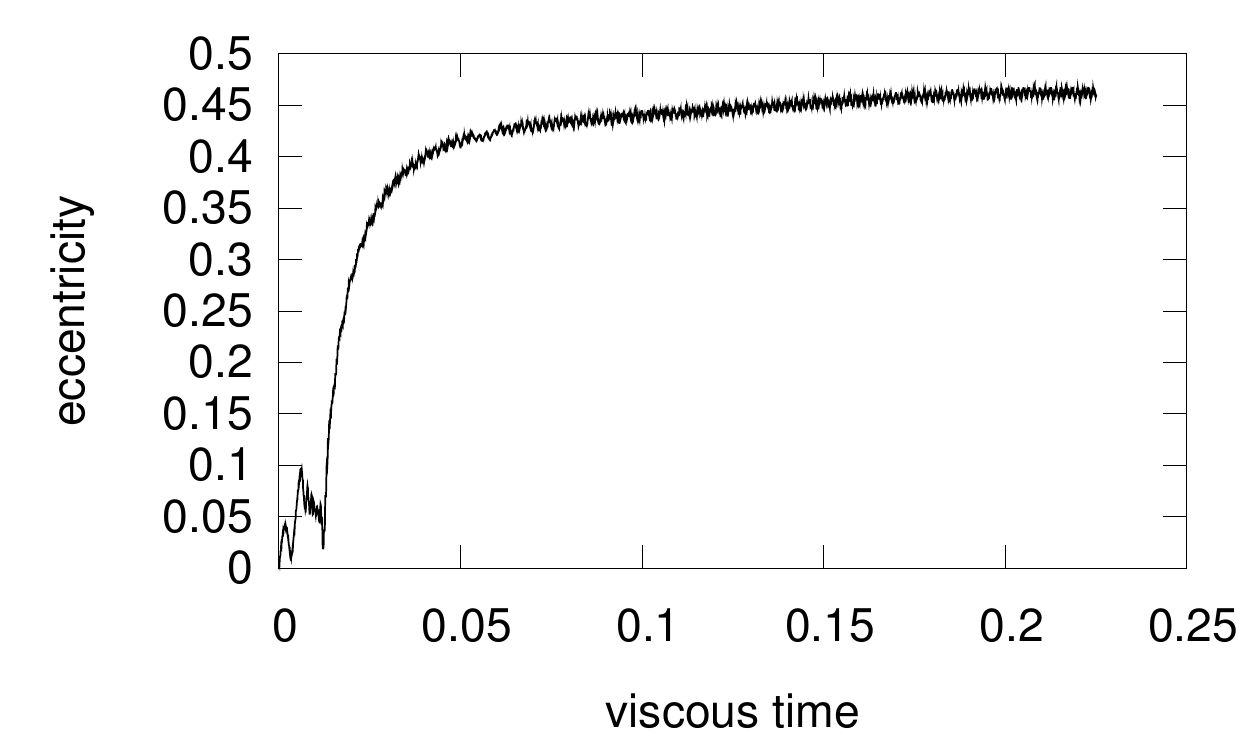}
\end{minipage}
\begin{minipage}{180mm}
\centering
\includegraphics[width=75mm]{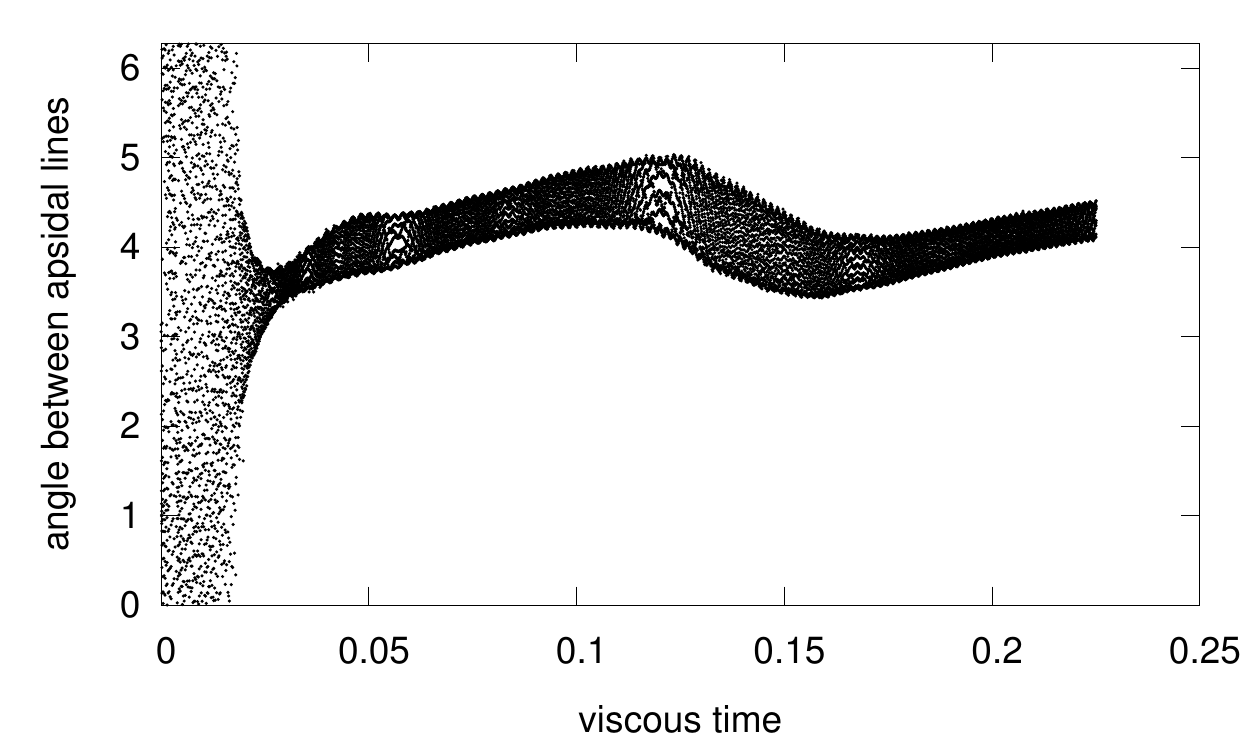}
\includegraphics[width=75mm]{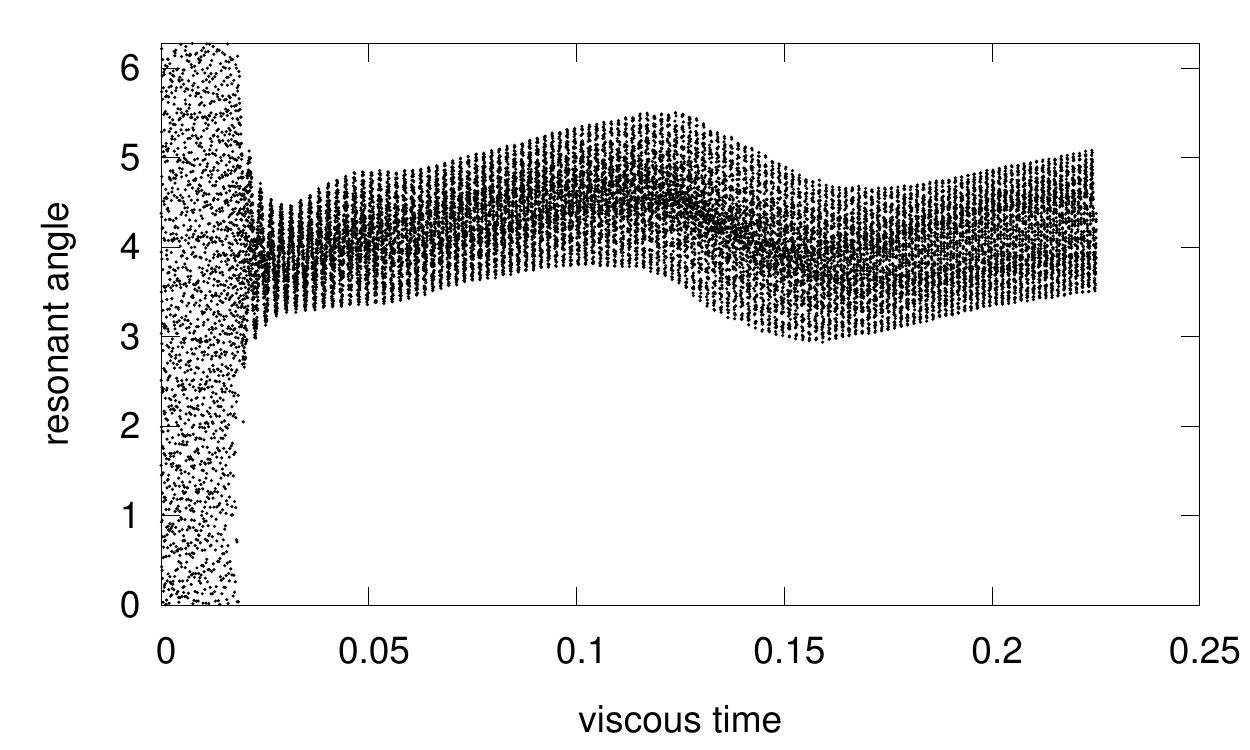}
\end{minipage}
\caption{\label{fig7}{
The evolution of the  semi-major axis ratios
(upper left panel),
the Super-Earth
eccentricities (upper right panel), the angle between the apsidal lines 
(bottom left
panel) and the resonant angle (bottom right panel) due to the disc-planet 
interaction for a discs with the
surface
density: $6\cdot 10^{-4}$
and the kinematic viscosity equal to $2\cdot 10^{-5}$.
}}
\end{figure*}

\subsection{Other configurations}
\label{other}
In the previous Subsection we have singled out within the parameter space
of typical physical properties of protoplanetary discs the
region which is favourable for the achievement of the 2:1 
resonance. We have also
characterized the final configurations
of the planets in this commensurability. However, the planet evolution  in 
gaseous discs does not always  end up in the 2:1 commensurability. 
In the following 
we would 
like to describe which are the alternatives to this behaviour
and 
to determine the disc 
properties that lead to these different possibilities. 

\begin{figure*}
\begin{minipage}{180mm}
\centering
\includegraphics[width=85mm]{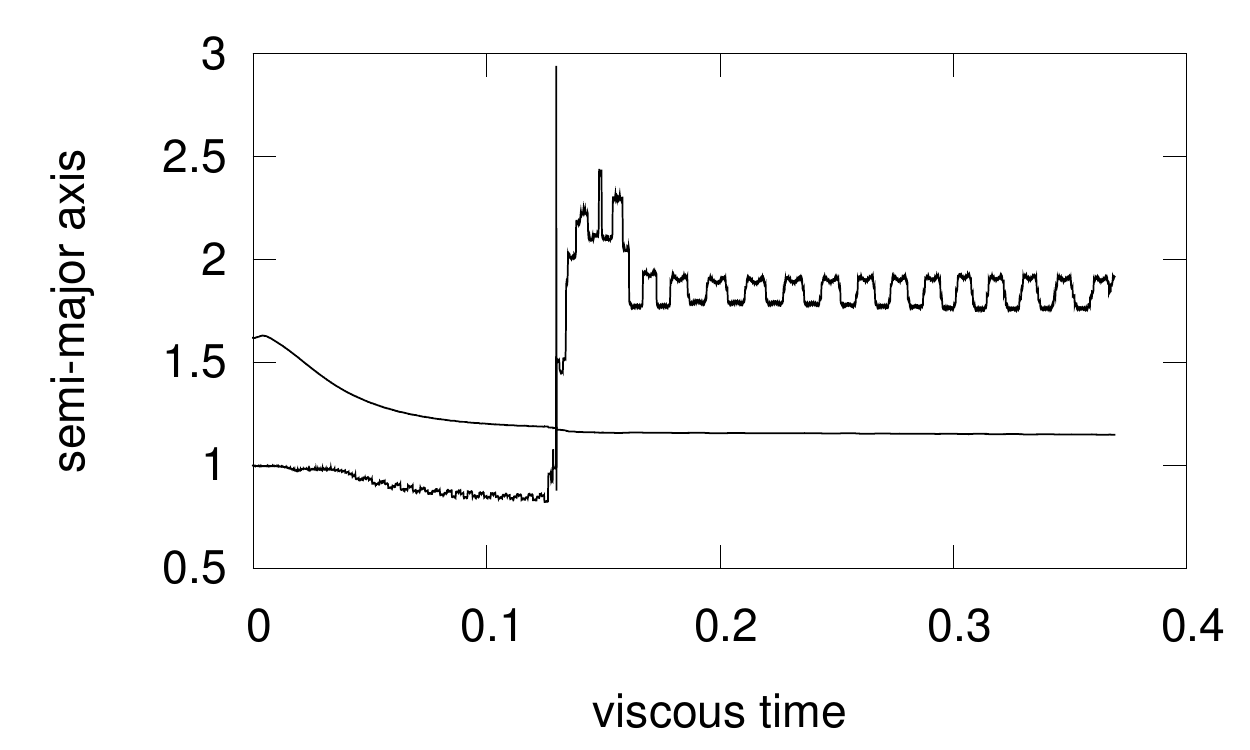}
\includegraphics[width=85mm]{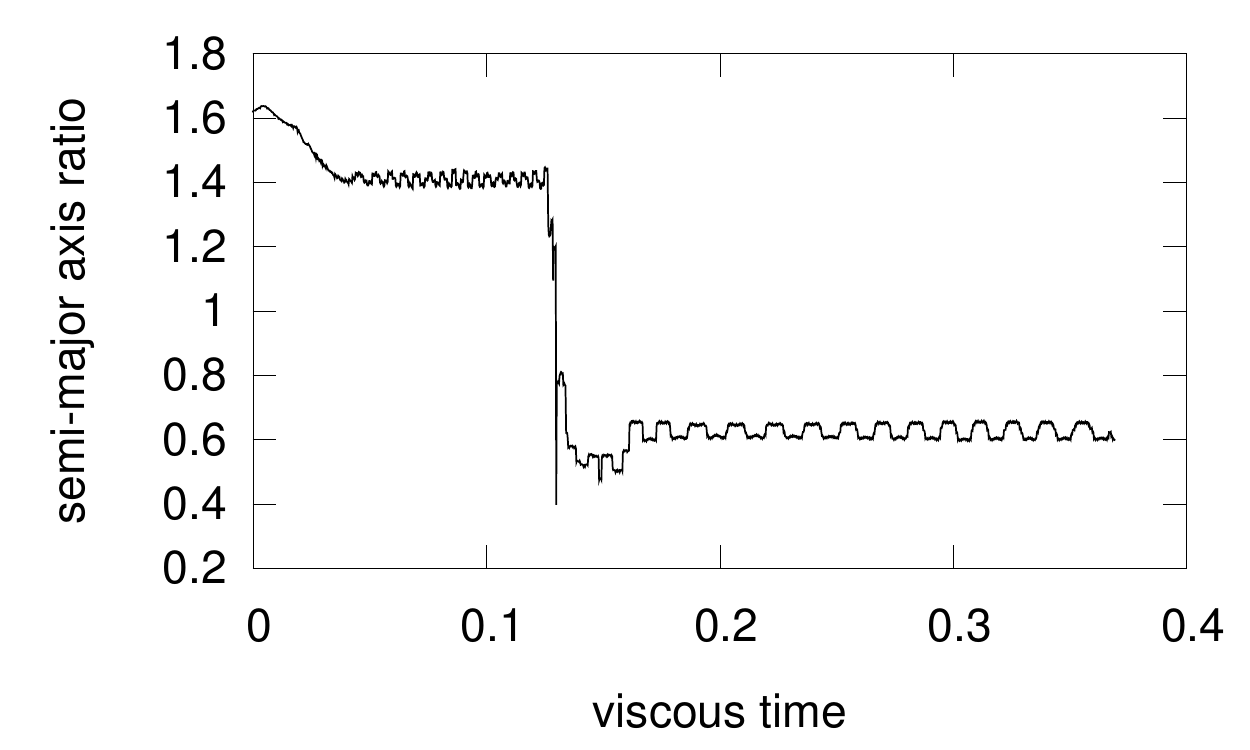}
\end{minipage}
\begin{minipage}{180mm}
\centering
\includegraphics[width=85mm]{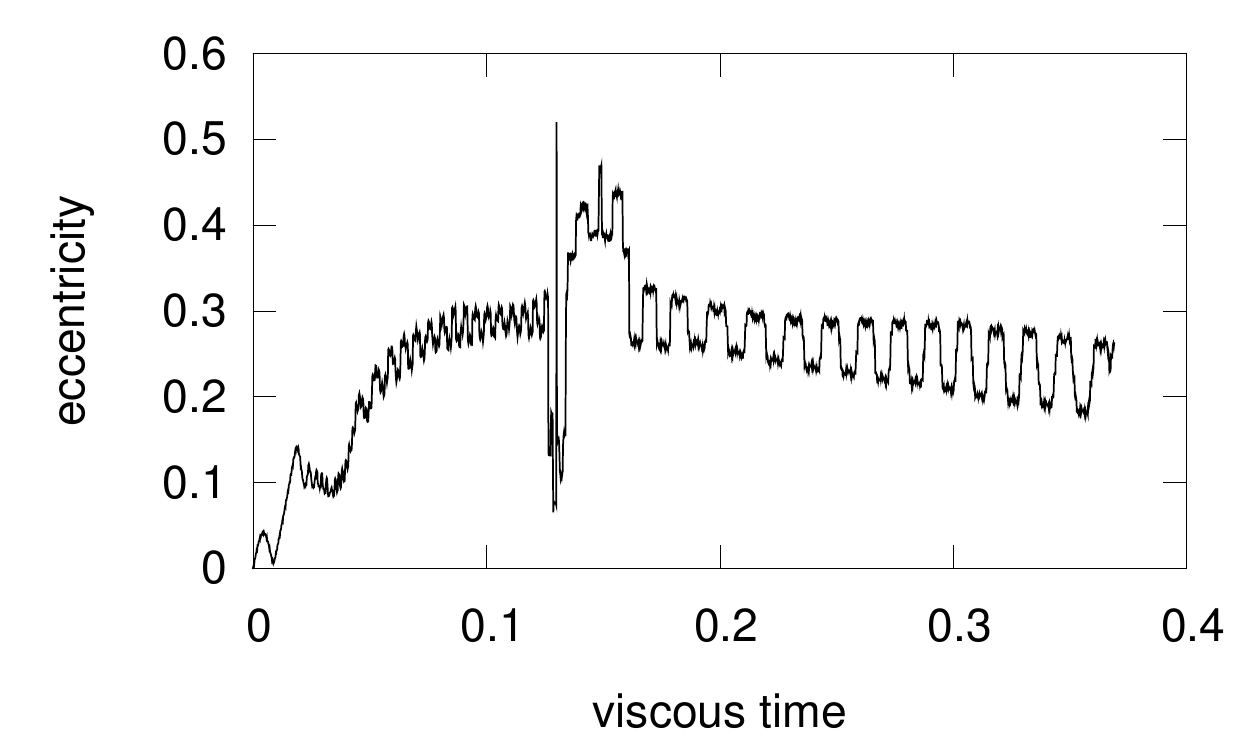}
\includegraphics[width=85mm]{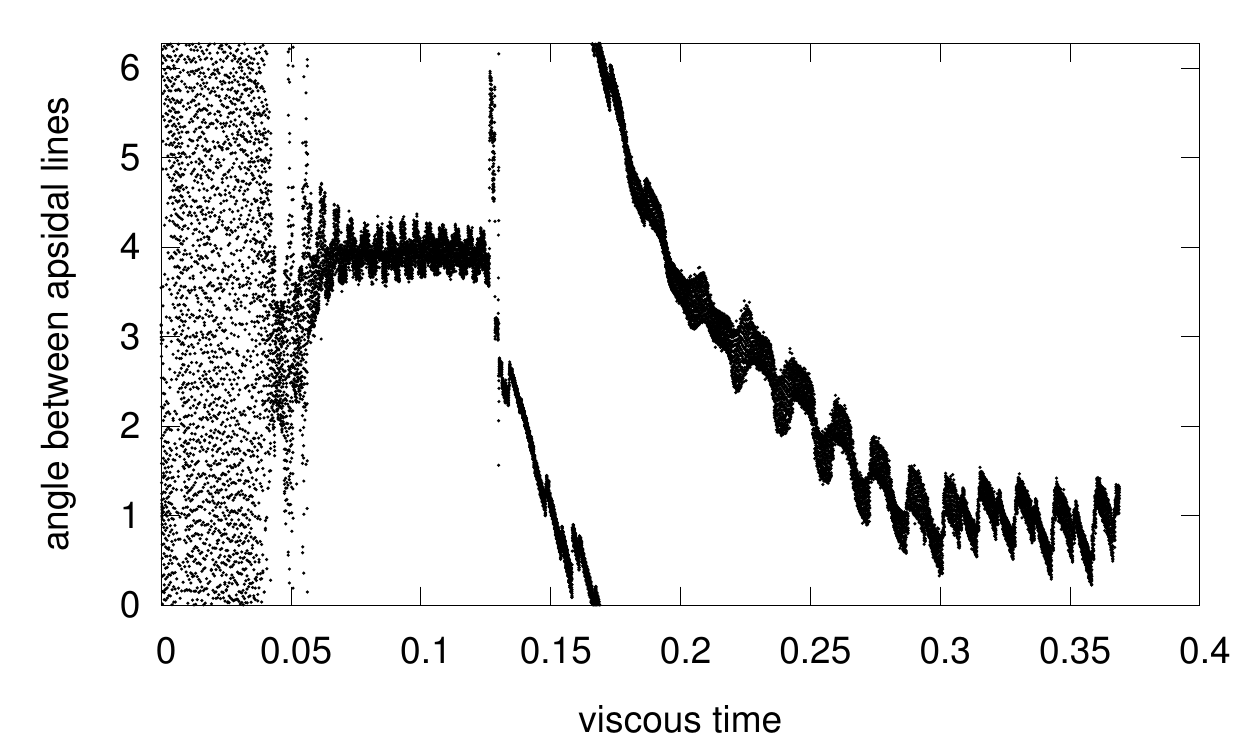}
\end{minipage}
\caption{\label{fig8}{An example of unstable second order commensurability
5:3, which ends up with orbit crossing and evolves into a configuration close
to the 1:2 resonance. The surface density is equal to $4.5 \cdot 10^{-4}$ 
 and the viscosity
has the value of $5 \cdot 10^{-5}$. The upper left panel shows the semi-major 
axis beheviour for both planets. The upper right panel illustrates the changes
of the semi-major axis ratio in time. The lower left panel displays the 
evolution of the eccentricity of the Super-Earth orbit. Finally, the lower 
right panel presents the angle between the apsidal lines.
}}
\end{figure*}

Let us start from the region in which the planets pass through 
the 2:1 resonance and
approach the 3:2 commensurability. This is the   small 
hatched region of Fig. \ref{fig1}
in which the surface densities vary between 
$4.5 \cdot 10^{-4}$ 
and $6 \cdot 10^{-4}$ in dimensionless units and the viscosities 
take their values in the
narrow range  
$2-3 \cdot 10^{-5}$.
For $\Sigma=6 \cdot 10^{-4}$ and $\nu=2 \cdot 10^{-5}$ 
the planets enter the
3:2 commensurability. 
The semi-major axis ratio oscillates around the value 
1.32 (Fig. \ref{fig7}).
The eccentricity of the Super-Earth reaches a value of 0.45 which is lower 
than what we get for the 2:1 commensurabilities. This result is in  good
agreement with that previously presented in \cite{paperI} for the 3:2
resonance.
 The angle $\Delta \omega$
between the apsidal lines  and the resonant angle librate around the 
value of 230 degrees (4 radians).  Usually one observes that
the equilibrium value of $\Delta \omega$ is either 0 or 180 degrees.
However,
it has been shown in \cite{beauge} 
that $\Delta \omega$ is described by 
a function of the masses and the eccentricities of the two 
planets and that, for high eccentricities, it can librate
around the values that are 
different from 0 or 180 degrees (e.g. \cite{kleyaps}).
In a disc with the same surface density ($\Sigma=6 \cdot 10^{-4}$)
but with  higher viscosity 
($\nu=3 \cdot 10^{-5}$), the 3:2 resonant trapping 
is also present, but it 
lasts only
for roughly 470 orbits and after that  the planets scatter. 
At the moment in which the planet scattering 
takes place, the eccentricity 
of the Super-Earth equals 0.38. The 3:2 commensurability has been achieved also
in a disc with the surface density $\Sigma=4.5 \cdot 10^{-4}$ and the 
viscosity $\nu=3 \cdot 10^{-5}$. In this case, the eccentricity of the Super-Earth 
is  equal to 0.3. The system does not abandon the 3:2 commensurability till
the end of our simulations.
For the same surface density $\Sigma=4.5 \cdot 10^{-4}$
and lower
viscosity ($\nu =2\cdot  10^{-5}$) we have  again scattering in the system, 
but  this time it occurs after staying for 
400 orbits  in the 3:2 resonance. 

Below the threshold of the viscosity necessary for the 2:1 resonant trapping,
which occurs at $4 \cdot 10^{-6}$, the migration is divergent, see the light
grey part in Fig. \ref{fig1}. This region of parameters does not contain
resonances, so we will not discuss it any further.

More interesting is the white region of high viscosities of Fig. \ref{fig1}.
In this part of the diagram we observe 
unstable behaviours leading to major changes in the planet configurations. 
One of the examples of those instabilities is shown
in Fig. \ref{fig8}. It is the case of a disc with  surface density
equal to $4.5 \cdot 10^{-4}$ and viscosity $5 \cdot 10^{-5}$. 
The gas giant evolving in the disc captures the Super-Earth in the 5:3
mean-motion resonance, as it is indicated by the fact that the
relevant resonant
angle is librating around a constant value (see lower right panel of
Fig. \ref{fig8}). 
At a certain time after the instant 0.1 in viscous time units
the low-mass planet starts  to go 
around its host star  
on a clearly chaotic orbit.
Further evolution leads to the orbit crossing
and the gas giant becomes the internal planet in the system. Moreover, the
final configuration after the orbit crossing is close to the 1:2 resonance. 

\section{Theoretical aspects of the resonance survey}
\label{theoretical}

In Fig. \ref{fig1} we have shown the behaviour of the planets in the
 surface density-viscosity plane obtained from  hydrodynamical 
simulations.
It might be useful to compare our numerical results with those obtained from
the analytical expressions for the migration speed provided in the literature. 
First, we assume that the Jupiter migrates according to the 
 type II migration (\cite{linpap93,crimor2007})  
 and its migration time can be estimated by
\begin{eqnarray}
\tau_{II}=\frac{(GM)^{1/2} m_J r_J^{1/2} }{2 \Gamma _{II}}
\label{torque}
\end{eqnarray}
where $r_J$ denotes the radial position of the gas giant,  
$m_J$ is the mass of 
the gas giant, and $\Gamma_{II}$ 
is the torque acting on the Jupiter. The torque is given by the 
following expression \citep{crimor2007}
\begin{equation}
\Gamma_{II} = 2\pi r^+ \Sigma_{LP}(r^+)v_r(r^+)\sqrt{GMr^+}
\label{cridawzor}
\end{equation} 
where
\begin{eqnarray*}
\Sigma_{LP}(r,t)=\Sigma_0
T^{-5/4}\Big ( \frac{\sqrt{r}-\sqrt{R_{inf}}}{\sqrt{r}}\Big ) 
\exp(-a(GMr)^2/T)
\end{eqnarray*}
is a surface density profile taken from \cite{lp}. Here, $T=12a\nu t+1$ 
is a
function of kinematic viscosity $\nu$ and time $t$,
$R_{inf}$ is the inner edge of the disc,  $\Sigma_0$ and $a$ are
constants defining the initial shape of the surface density distribution. 
The radial velocity $v_r$ as well as all other quantities in the expression
for the torque acting on the planet is evaluated at the external
edge of the gap opened by the gas giant, namely at
$r^+ = r_J+2R_H$ where 
$R_H$ is the radius of the Hill sphere.
Having specified the migration time for the gas giant we turn our 
attention to the Super-Earth.
The evolution of the Super-Earth proceeds according to the type 
I migration as described  in
\cite{tanaka02}
\begin{eqnarray}
\tau_{I}=(2.7+1.1p)^{-1}\frac{M}{m_s}\frac{M}{\Sigma_s
{r_s}^2}h^2{\Omega_s}^{-1}
\label{tanakawzor}
\end{eqnarray}
where p denotes the surface density slope of the disc (see Eq.~(\ref{sigma})), 
$m_{s}$ is the mass of the Super-Earth and $\Omega_{s}$ is the angular velocity
at the Super-Earth location $r_{s}$.
 Thus, we can easily predict when the differential migration 
will be convergent, which is essential for the occurrence of the 
mean-motion resonance. 
From the Eqs.~(\ref{cridawzor}) and (\ref{tanakawzor}) the criterion for 
the convergent 
migration reads
\begin{eqnarray}
\nu & > & {(1.364+0.541p) \over 3\pi} \times
\nonumber \\
& &\frac{ \left( {r_J \over r^+}\right)^{1/2}
\left( {m_{s} \over M}\right) \left( {m_J \over M}\right)\left( 
1- \left({R_{inf} \over {r^+}} \right)^{1/2}\right)}
{h^2\left( 1- {4a \over T}(GMr^+)^2 \left( 1- \left({R_{inf} \over r^+} 
\right)^{1/2}\right) \right) }
\label{criterion}
\end{eqnarray} 
Moreover, we can  combine the expected 
differential migration speed with the 
strength of 
the 2:1 and 3:2 commensurabilities given by \cite{alice}, \cite{mark} 
in order to 
guess if
the capture will take place or not. 

\begin{figure*}
\begin{minipage}{180mm}
\centering
\includegraphics[width=85mm]{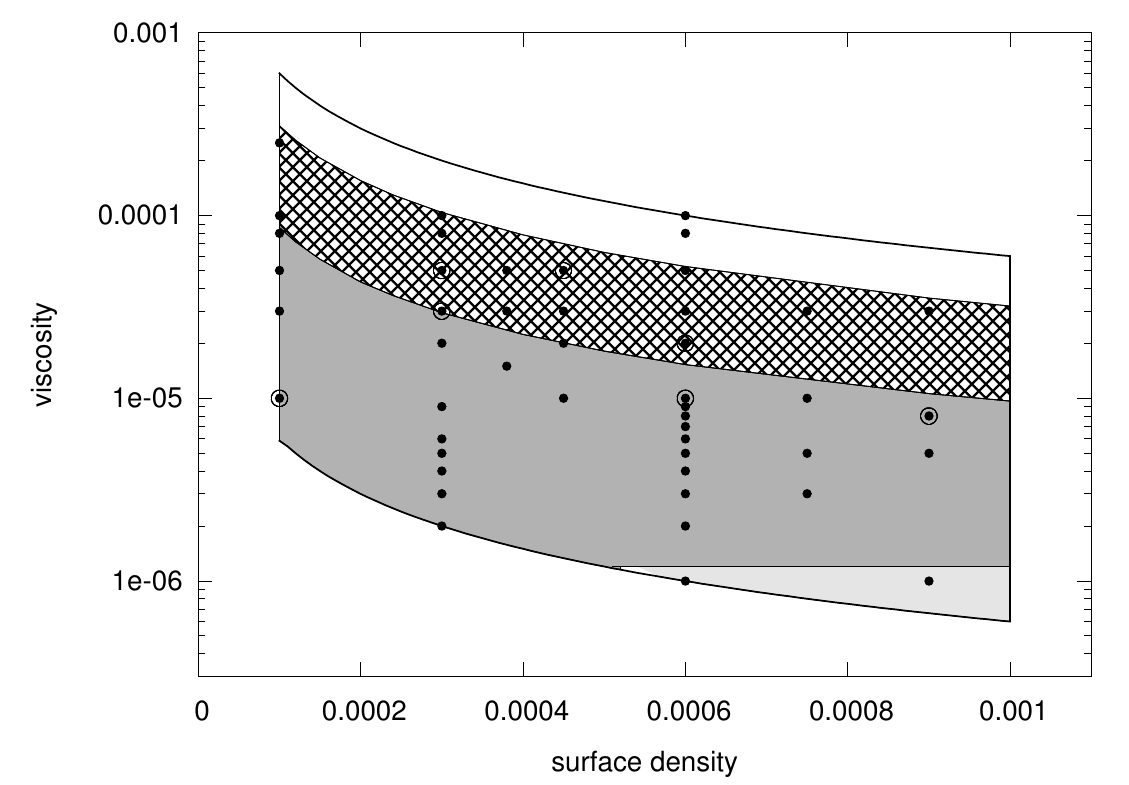}
\includegraphics[width=85mm]{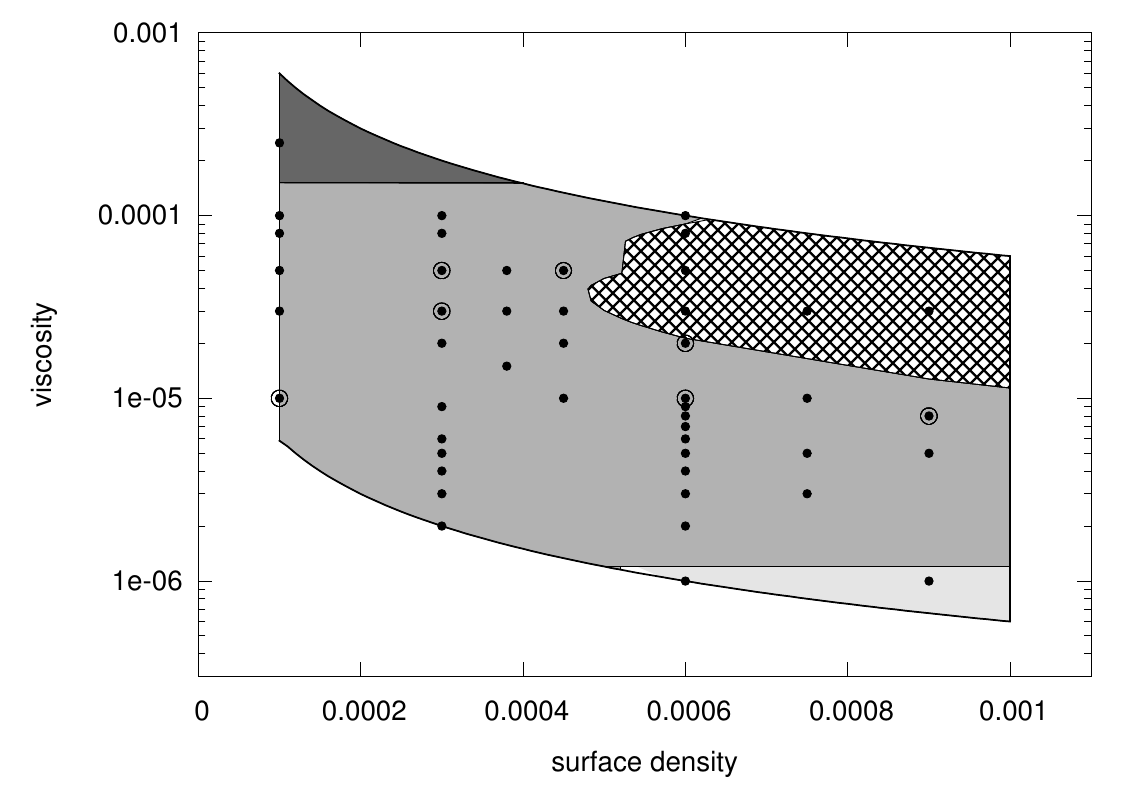}
\caption{\label{fig11}{The parameter space of the properties of the
protoplanetary discs: surface density - viscosity plane obtained
under the assump\-tions that the planets evolve according to
the classical type I
and type II migration respectively (left panel) and that the evolution
of the Super-Earth is the type I migration and that of the Jupiter is 
modified by the gas  possibly present in the gap
\citep{crimor2007} (right panel).	
}}
\end{minipage}
\end{figure*}

The resonant capture for the first order resonances in the restricted 
three body problem occurs when
\begin{eqnarray}
\frac{1}{\frac{1}{\tau_I}-\frac{1}{\tau_{II}}} \geq
\frac{3 \pi}{ \dot{\eta}_{crit} \Omega} 
\label{alicewzor}
\end{eqnarray}
where 
$\dot \eta _{crit}$ is the critical
mean motion drift rate and $\Omega$ is the angular velocity of the Jupiter.
In the case of internal 2:1 resonance
$\dot{\eta}_{crit}=22.7(m_J/M)^{4/3}$ and for
the 3:2 commensurability $\dot{\eta}_{crit}=126.4(m_J/M)^{4/3}$ \citep{alice}. 
Finally, we  make an attempt to
draw the theoretical analog of our Fig. \ref{fig1}. The results are shown 
in
Fig. \ref{fig11} (left panel). 
In this figure we have used the same grey scale and patterns  as in 
Fig. \ref{fig1} in order to facilitate the comparison.  

The overall theoretical picture is similar to the one obtained in our
numerical simulations. However, the details do not match exactly. The
convergent migration takes place for viscosities higher than  
$1.2 \cdot 10^{-6}$. This gives the region of the divergent migration
smaller than in Fig. \ref{fig1}. 
The region of the 2:1 resonance occurrence
in both numerical and theoretical approaches occupies almost the same  
place in the space of the disc parameters. The only significant
differences are present for relatively low surface densities and high 
viscosities. The hatched region with high viscosities (see  
Fig.  \ref{fig11}, left panel) for which there are the conditions 
for the 3:2 commensurability, is much more extended than the analogous
region that results from our numerical simulations.  In the white region we
expect the 4:3 commensurability. 
However, 
we would like to stress here that the picture discussed
above has been obtained under the assumptions that the planets evolve 
according to classical type I and type II migrations. 

In a more realistic
situation, when the gap opened by the gas giant is not deep and clean
the torque acting on the planet needs to be modified (\cite{crimor2007}).
Therefore, in a more refined approximation, we have changed the assumption about gas giant
migration. Namely, instead of applying Eq.(10) of \cite{crimor2007} as we
did in order to obtain the diagram displayed in Fig.  \ref{fig11} (left
panel) we use Eq.(15) of that work, so now the torque reads
\begin{eqnarray*}
&\Gamma & =2\pi r^+ \Sigma_{LP}(r^+)v_r(r^+)\sqrt{GMr^+} \times 
\nonumber \\
& &\left[ 1-f(P)-\frac{10R_H}{\pi r^+}\frac{\Omega_J
{r_J}^2}{\sqrt{GMr^+}}\frac{v_r(r_J)}{v_{r}(r^+)}f(P)
\frac{d \log\Sigma/B}{d \log
r} \right]
\label{full}
\end{eqnarray*}

\hspace{7.4cm} (9) \\
 where 
$\Sigma$ is the density inside the gap, $B$ is the second Oort constant. 
The function \[ f(P) = \left\{
\begin{array}{ll}
(P-0.541)/4 \  \mbox {\ \ \ \ \ \ if  $P<2.4646$ }
\\
\\
1-\exp(-P^{0.75}/3) \  \mbox{ if $ P \geq 2.4646$} \\
\end{array}
\right . \] describes the gap depth expressed as a ratio of the gap
surface density to the unperturbed density at $r^+$
and $P$ is defined by 
\begin{eqnarray*}
P=\frac{3H}{4R_H}+\frac{50}{(m_J/M) R} \ltsimeq  1
\end{eqnarray*}
where $R$ is the Reynolds number. 
In this way we are able to take into account the torque
exerted on the outer disc by the gas in the gap and the 
corotation torque.
The resulting picture can be viewed in Fig.  \ref{fig11} (right panel). 
As it is possible to see in this figure the hatched area has shrunk
towards higher values of the surface density, which is in better 
agreement with the numerical findings. 
 The region of the divergent migration for low viscosities
remains unchanged. This is because, for low viscosities, the torque 
expressed by Eq. (\ref{cridawzor}) is very similar to that of Eq. (\ref{full}),
so
the line dividing
the regions of the divergent and convergent migrations can be obtained 
from Eq. (\ref{criterion}) also in this case. Moreover, the region 
of the
divergent migration has appeared for high viscosities as expected.
 Whenever the migration is convergent, the planets will be captured 
either in the 2:1 or 3:2 resonances. The line between the regions of the
occurrence of these 
commensurabilities can be found using the criterion Eq. (\ref{alicewzor})
in which $\tau_{II}$ is expressed by  Eq. (\ref{torque}), but  
$\Gamma_{II}$ is replaced by the torque given by
Eq. (\ref{full}). 
The stability of the resonances requires further studies.
Other methods must be used in order to determine the final outcome of the
migration of the resonant pair of planets. The numerical approach demonstrates
to be very useful to tackle such problems and enables us to identify those
regions in the parameter space in which the commensurabilities are not
only form, but also are maintained at least for the time accessible to
full hydrodynamic calculations. In Fig.  \ref{fig1} those regions has been
indicated by the medium grey colour and the hatched pattern. The rest of the
parameter space, where the migration of planet is convergent has been
left in white. The whole variety of behaviours has been observed in this white
region
including  transient second order resonances and the orbit crossing, which
indicates that not all resonant captures shown in Fig. \ref{fig11} 
are stable.

As already explained,
in order to draw Fig. \ref{fig11} we have used the assumption 
that the 
Super-Earth migrates according to \cite{tanaka02}, but recently
it has been shown by \cite{PaarMel2006}, \cite{BarMas2008}, 
\cite{KleyCrida2008}, \cite{Paa2010a, Paa2010b}
that in non-locally isothermal discs the rate and even the direction of
the type I migration can be strongly changed. 
Therefore, we discuss briefly  how our  results would be affected by
considering non-locally isothermal discs. If the Super-Earth
 migrates inward slower than in our simulations
 we would observe a faster differential migration and in 
consequence the regions of occurrence of the 2:1 resonance
 in our surface density-viscosity plane
 would be shifted downward.  
In the situation, when the Super-Earth migrates inward
faster than in our calculations, it 
would be more difficult to obtain the convergent migration. 
Thus, the regions of occurrence of the 2:1 resonance
 would be shifted upward. However, we expect that the global 
picture should be similar. In the case of the Super-Earth migrating outwards, 
the most likely effect would be 
the scattering
of the low-mass planet.

We have  also
checked the influence of the surface density profile on our final results.
 In our calculations we have used the initial density profile with
 power-law index $p=0$, while the observations suggest $0.5<p<1.5$ 
\citep{pietu2007}. In doing that, we have run simulations with $p=1$ 
taking
$\Sigma=6 \cdot 10^{-4}$ and  viscosities $\nu= 10^{-6}, 10^{-5},
 2 \cdot 10^{-5}, 3 \cdot 10^{-5}$ and $8 \cdot 10^{-5}$. 
It has been found that the differential migration is slower than for
 $p=0$, 
so that the global picture is qualitatively the same but the regions 
of the characteristic behaviours in Fig. \ref{fig1} are moved upward.
 For example the case with $\Sigma=6 \cdot 10^{-4}$
  and $\nu=2 \cdot 10^{-5}$ gives the 3:2 resonance 
for the disc with $p=0$. In the same disc with power-law index $p=1$
we have obtained the 2:1 commensurability.

Moreover, as we have mentioned in Section \ref{survey},
we do not exclude any material from the Hill sphere.
To justify our approach, we have performed 
numerical simulations in order to investigate 
the influence of the material close to
the planet on the gas giant migration rate. 
We have observed that excluding the material from the Hill 
sphere results in the
slowing down of the migration of the Jupiter, which has been already shown by
\cite{crida09}. Thus, the differential 
migration is also slowed down and the global picture would be the 
same as taking higher $p$. That means that the region of the resonant behavior
will be shifted toward higher viscosities in our parameter space.

\section{Observational aspects of the resonance survey}
\label{observational}

In our calculations we have started the investigations by choosing 
the masses of the planets and  setting the main properties of 
the protoplanetary disc in which they are embedded, namely its surface 
density distribution,  aspect ratio and viscosity. Ideally, one
would like to use  observed quantities in order to fix the initial conditions
for such calculations. However,  regardless of the fact that
circumstellar discs are  the natural outcome of the star-formation process,
their structure is still not known in sufficient details in order to set
such conditions using direct measurements.

\begin{figure*}
\begin{minipage}{180mm}
\centering
\includegraphics[width=120mm]{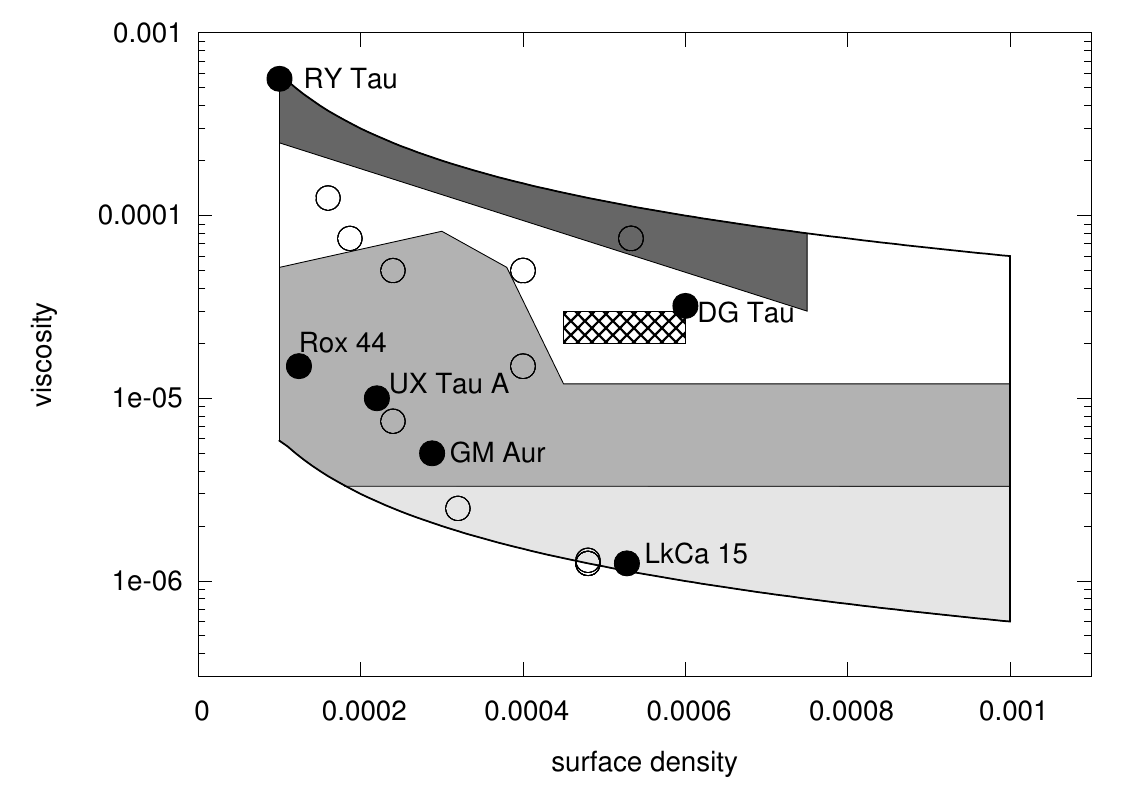}
\caption{\label{fig10}{The surface density - viscosity plane with the
positions of several interesting protoplanetary discs. See text for full
description.}}
\end{minipage}
\end{figure*}

The most powerful technique which could be able to  change this situation and 
provide the most common properties of  protoplanetary discs, is that of  high 
resolution imaging.
Recently, \cite{isella} have presented the CARMA (Combined Array for Research
in Millimeter-wave Astronomy) observations of the thermal dust emission
from the discs around RY Tau and DG Tau with the best resolution achieved 
to date at wavelengths of 1.3 mm and 2.8 mm, namely
0.15" (which gives 20 AU at the distance of the Taurus cloud). 
On the basis of their high angular resolution observations, \cite{isella} 
have determined the disc properties of these objects. Unfortunately, at 
radii smaller than 15 AU, which are the most relevant for our studies discussed
here,
the surface density is model dependent and can 
vary by almost an order of magnitude. 
In our work, as  mentioned in Section \ref{survey}, 
we simply assume 
the classical power law parameterization of Eq. \ref{sigma} with $p=0$.
A better assumption for the surface density distribution will be possible 
only with
the future advances in observational techniques. Taking the values of the
surface density derived by Isella et al. (2010) we have placed DG Tau and 
RY Tau in the diagram of Fig. \ref{fig1}. The result is displayed in 
Fig. \ref{fig10} which shows the same regions as Fig. \ref{fig1}. 
DG Tau  is located in that region of the parameter 
space in which a 3:2 resonance might be achieved but it is not preserved.
In the case in which  planets would be present in DG Tau,
the most likely scenario would be the temporary Super-Earth capture 
into 3:2 commensurability by the Jupiter-like planet and then scattering. 
This case has been described shortly
in Section \ref{other}. Also for the RY Tau we do not expect the 2:1 
commensurability. 
If in the inner disc of RY Tau there would be
 a pair of planets similar to that considered in our 
paper, then  
the Jupiter-like planet would migrate outwards
and the Super-Earth inwards. The divergent migration would not induce
any commensurability. 

We have also collected from the literature the observed properties of the
circumstellar discs around other young stars. In Fig. \ref{fig10}
we have 
marked the location of several observed systems. 
Rox 44, UX Tau A and LkCa 15 have been observed and 
modelled by \cite{espaillat}. The data has been taken from their Table 3.
The physical properties of Rox 44 and UX Tau A indicate the possibility of
the occurrence of the 2:1 commensurability. This is also the case
of GM Aur studied by \cite{hughes}. Instead, the disc in 
LkCa 15 is located well
below the region where the resonant structure can be expected. If planets
were  present in this  disc,
the gas giant would migrate slower than the Super Earth and, since the
Jupiter is 
on the external orbit, it would not be  able to catch the Super-Earth and form
the mean-motion resonance.   
The open circles are the locations of ten other discs studied by 
\cite{Andrews}, namely Elias 24, GSS 39, DoAr 25, WAOph 6, VSSG 1, SR 4, SR 13,
WSB 52, WL 18 and SR 24 S. They belong to the 1 Myr-old Ophiuchus 
star forming region. Two of those, namely DoAr 25 and SR 13,
share  almost the same properties and  are located very close
to each other in our picture of the parameter space.

\begin{figure*}
\begin{minipage}{180mm}
\centering
\includegraphics[width=85mm]{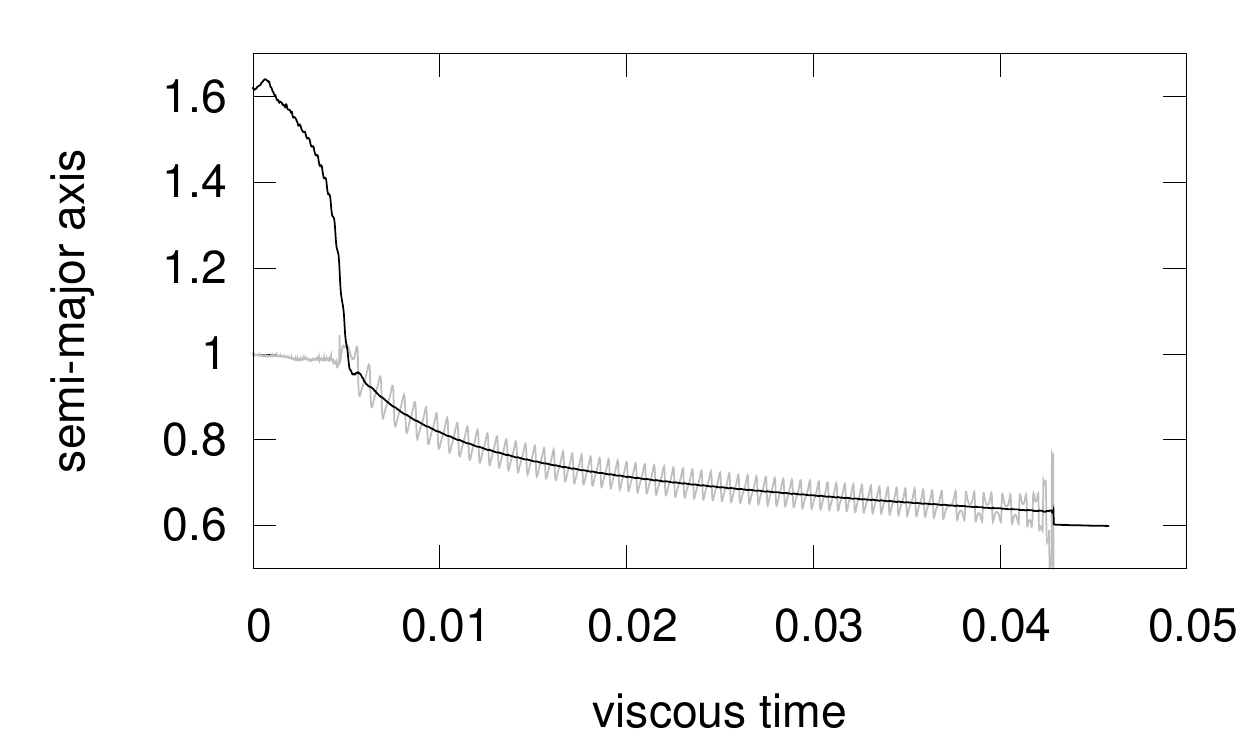}
\includegraphics[width=85mm]{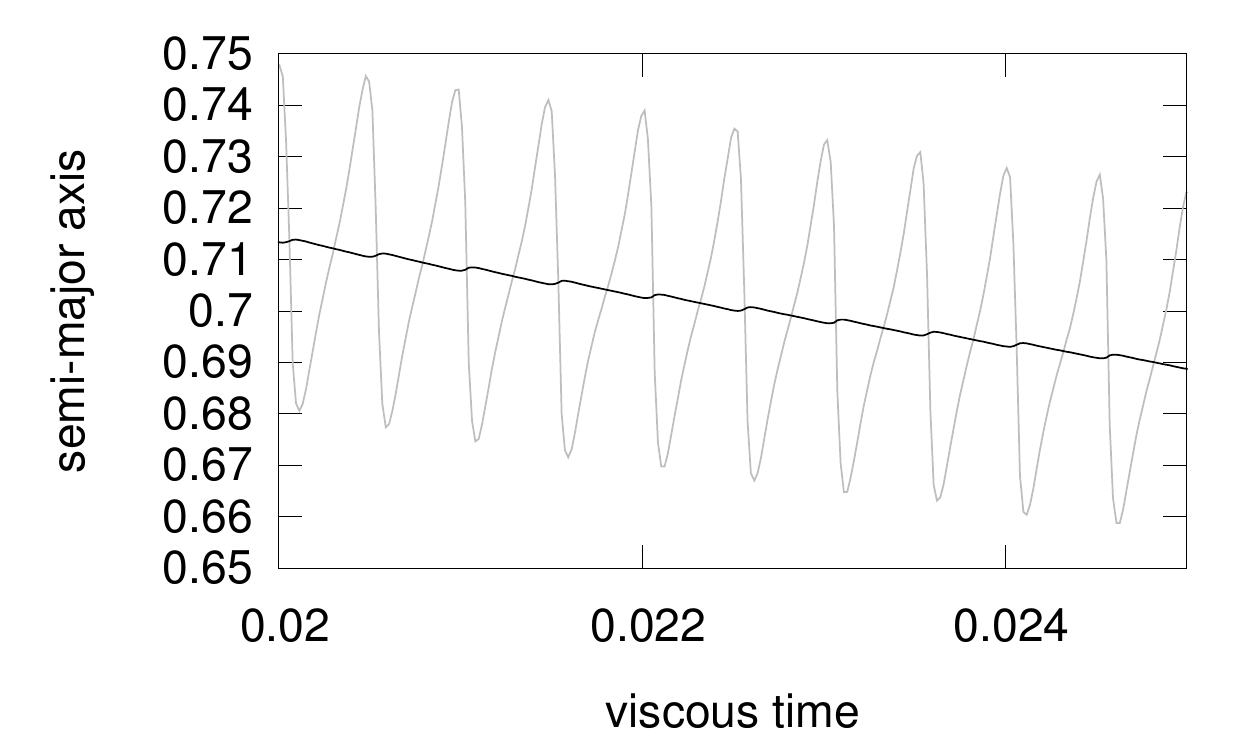}
\end{minipage}
\begin{minipage}{180mm}
\centering
\includegraphics[width=85mm]{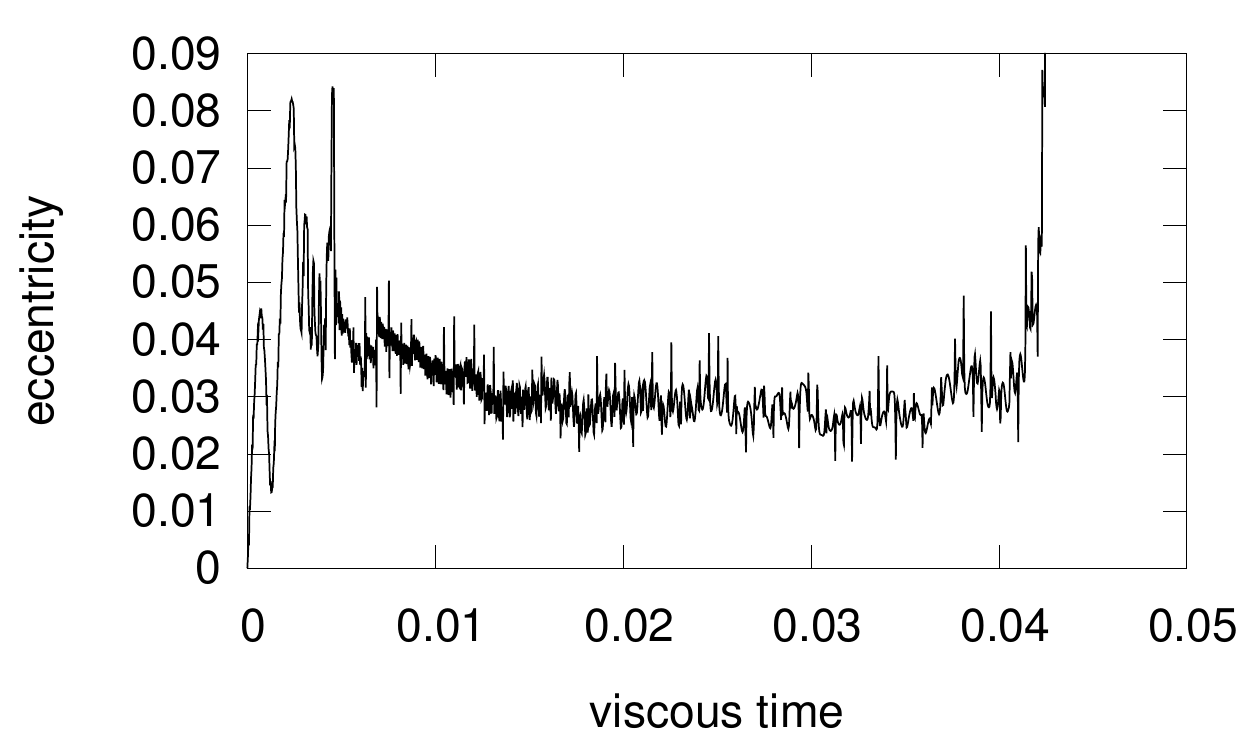}
\includegraphics[width=85mm]{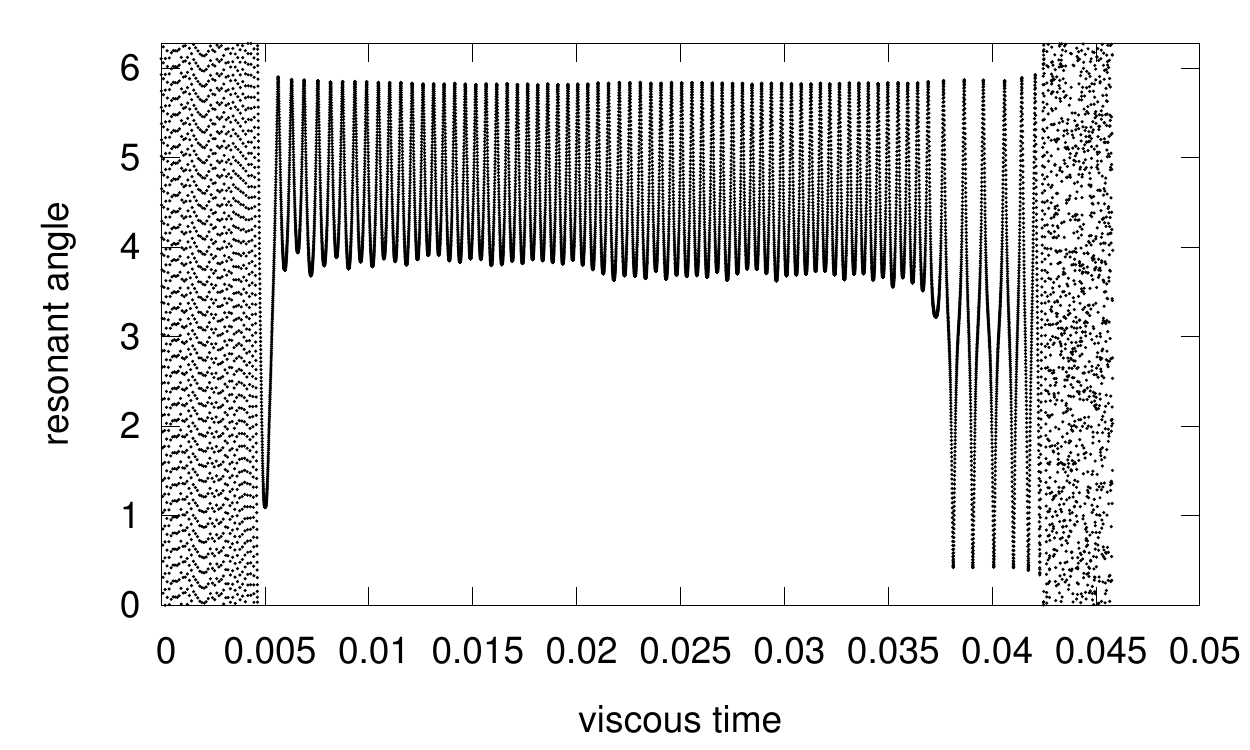}
\end{minipage}
\caption{\label{fig9}{An example of the 1:1 commensurability,
which ends with the ejection of the Super-Earth from the system.
The two upper panels show the semi-major axis evolution for both planets.
The upper right panel is a zoomed out version of the short time interval 
presented in the upper left panel. The lower left panel presents how the
Super-Earth eccentricity changes in time and lower right panel illustrates
the behaviour of the resonant angle.}}
\end{figure*}

Of course, the predictions coming from the observations are still somewhat 
vague due to the huge uncertainties in the determination of the disc
structure parameters. The purpose of this Section was however to stress the
fact that
such predictions are already possible and 
can be tested in the future 
with the advances of the observational
techniques.

\section{Discussion and conclusions}
\label{conclusions}
In this paper we have constrained the properties of
protoplanetary discs which favour the locking of a Super-Earth
and a Jupiter-like planet orbiting around a Sun-like star
in the 2:1 commensurability.
The Super-Earth and the Jupiter are  located on the internal 
 and external orbits respectively.
The results are illustrated in 
Figs. \ref{fig1} and \ref{fig10}.
In both figures the medium grey region is where we expect the 2:1
resonance to occur.
In the low surface density regime (less or
equal to $4\cdot 10^{-4}$) the final outcome is relatively easy to determine. 
The 
migration of both planets is rather slow in this region. We expect the 2:1 
resonance 
for viscosities not lower than $4\cdot 10^{-6}$. 
Below this threshold
the Jupiter 
migrates slower than the Super-Earth and no commensurability can be achieved.
On the other side,
in order to get the 2:1 resonance, the viscosity cannot be too high,
because then the Jupiter will migrate outwards. 
This is the case in which the viscosity 
combined with the surface density through Eq. (\ref{signu}) gives
accretion rates higher than $10^{-7}$ M$_{\odot}$/yr. 
In the high surface density regime (higher than $6\cdot 10^{-4}$) 
the situation is more
complicated  and the 2:1 resonance is allowed only in a narrow strip where
the viscosity is varying in the range from 
$4 \cdot 10^{-6}$ till $10^{-5}$.
In this regime  the equilibrium gap profile is not empty. 
This is particularly true for 
relatively high viscosities. For that reason, putting the gas giant in 
the disc results in  its very rapid migration. To
avoid the artefact caused by the process of gap opening, we kept the 
Jupiter-like planet
fixed, not allowing for its migration till the gap has reached the
quasi-stationary state and stops changing significantly. In this way we 
have obtained the 2:1 resonance for viscosities ranging from  
$4\cdot 10^{-6}$ till $10^{-5}$. The results coming from numerical simulations
have been compared with the analytical predictions summarized in 
Fig. \ref{fig11} and with the current observations, see Fig. \ref{fig10}.

The orbital migration of planets embedded in  protoplanetary discs is a
natural mechanism in which  resonant configurations might occur. 
However, it is not   
the only way to form  orbital commensurabilities. It has been
suggested that the planet-planet scattering might be an efficient mechanism
leading to the resonant capture of planets \citep{raymond}. 
Our computational set up allows us to investigate also this scenario in the
situation when the gaseous disc is still present in the system. 
We have employed a simple procedure to locate both planets in a
disc with high surface density (higher than $6\cdot 10^{-4}$) and 
let them migrate
without waiting for the Jupiter to open a gap. Here we discuss  in details
one example of resonance obtained in this way.
Our probe, which consists of a Jupiter-like planet
and a Super-Earth, is located in a disc with  surface density equal to
$9\cdot 10^{-4}$ and  viscosity value taken to be 8$\cdot 10^{-6}$. 
As expected, 
the Jupiter
migrates extremely fast  (with type III migration, \cite{massetpap}), 
traveling one third of the distance from its initial
location to the star in just 100 orbits. Its orbit crosses the orbit of 
the slowly
migrating Super-Earth and we witness a resonant capture into 1:1 mean motion
resonance. This is shown in Fig. \ref{fig9} (two upper panels). After the 
occurrence of the locking the Super-Earth stays very close to the Jupiter on 
the horseshoe orbit
with an eccentricity of about 0.03 (Fig. \ref{fig9}, left lower panel). The 
resonant angle
is shown in Fig. \ref{fig9} 
(right lower panel). 
The 1:1 commensurability created in the way described above lasted 
for roughly 700 orbits and after that the Super-Earth has been
ejected from the system. 

Concluding, we have shown how  the
mean-motion resonances observed in the already formed planetary systems
can help in getting additional insight into the conditions occurring in  
protoplanetary discs in the early phases of the planetary system evolution. 
In other words, we  have made an attempt to  constrain the resonant 
properties of the  planetary systems which may occur in observed 
protoplanetary discs.
Assuming that the resonances
are due to early stages of the evolution, we can answer the question
on what should be the disc properties in order to attain  the 2:1 resonance.

\section*{Acknowledgments}
This work has been partially supported by MNiSW grant N203 026 32/3831
(2007-2010) and  MNiSW PMN grant - ASTROSIM-PL ''Computational Astrophysics.
The formation  and evolution of structures in the universe: from planets to
galaxies'' (2008-2010).
We acknowledge support from the Isaac Newton Institute programme ``Dynamics
of Discs and Planets''.
Part of this research was performed during a stay  at the Kavli 
Institute for
Theoretical Physics, and was supported in part by
the National Science Foundation under Grant No.
PHY05-51164.
The simulations reported here were performed using 
the HAL9000
cluster of the Faculty of Mathematics and Physics of the University of
Szczecin. 
We thank the referee for valuable comments
which helped us in improving the manuscript.
We are grateful to John Papaloizou for enlightening discussions.
We wish also to thank Adam {\L}acny for his helpful
comments.
Finally, we are indebted to
Franco Ferrari for his continuous support in the development of our
computational
techniques and computer facilities.



\label{lastpage}

\end{document}